\documentclass[times]{qjrms4}

\usepackage{natbib}
\usepackage{epsfig}\usepackage{graphicx}
\usepackage{amsmath,amssymb,amsfonts}

\usepackage{booktabs}

\newcommand{\peq}{p_{{\rm eq}}}
\newcommand{\comment}[1]{}

\makeatletter
\newcommand{\xleftrightarrow}[2][]{\ext@arrow 3359\leftrightarrowfill@{#1}{#2}}
\newcommand{\xdashrightarrow}[2][]{\ext@arrow 0359\rightarrowfill@@{#1}{#2}}
\newcommand{\xdashleftarrow}[2][]{\ext@arrow 3095\leftarrowfill@@{#1}{#2}}
\newcommand{\xdashleftrightarrow}[2][]{\ext@arrow 3359\leftrightarrowfill@@{#1}{#2}}
\def\rightarrowfill@@{\arrowfill@@\relax\relbar\rightarrow}
\def\leftarrowfill@@{\arrowfill@@\leftarrow\relbar\relax}
\def\leftrightarrowfill@@{\arrowfill@@\leftarrow\relbar\rightarrow}
\def\arrowfill@@#1#2#3#4{%
  $\m@th\thickmuskip0mu\medmuskip\thickmuskip\thinmuskip\thickmuskip
   \relax#4#1
   \xleaders\hbox{$#4#2$}\hfill
   #3$%
}
\makeatother

\begin{document}
\runningheads{T. Berry and J. Harlim}{Semiparametric forecasting and filtering}
\title{Semiparametric forecasting and filtering: correcting low-dimensional model error in parametric models}
\author{Tyrus Berry,\affil{a}\corrauth\
John Harlim\affil{a,b}}

\address{\affilnum{a}Department of Mathematics, the Pennsylvania State University, 109 McAllister Building, University Park, PA 16802-6400, USA \\
\affilnum{b}Department of Meteorology, the Pennsylvania State University, 503 Walker Building, University Park, PA 16802-5013, USA}

\corraddr{Department of Mathematics, the Pennsylvania State University, 109 McAllister Building, University Park, PA 16802-6400, USA. E-mail: tyrus.berry@gmail.com}

\begin{abstract}
Semiparametric forecasting and filtering are introduced as a method of addressing model errors arising from unresolved physical phenomena.  While traditional parametric models are able to learn high-dimensional systems from small data sets, their rigid parametric structure makes them vulnerable to model error.  On the other hand, nonparametric models have a very flexible structure, but they suffer from the curse-of-dimensionality and are not practical for high-dimensional systems.  The semiparametric approach loosens the structure of a parametric model by fitting a data-driven nonparametric model for the parameters. Given a parametric dynamical model and a noisy data set of historical observations, an adaptive Kalman filter is used to extract a time-series of the parameter values.  A nonparametric forecasting model for the parameters is built by projecting the discrete shift map onto a data-driven basis of smooth functions.  Existing techniques for filtering and forecasting algorithms extend naturally to the semiparametric model which can effectively compensate for model error, with forecasting skill approaching that of the perfect model. Semiparametric forecasting and filtering are a generalization of statistical semiparametric models to time-dependent distributions evolving under dynamical systems.  
\end{abstract}


\maketitle

\section{Introduction}

The backbone of our modern numerical weather prediction (NWP) is the probabilistic point of view that was taken by \cite{epstein:69} and \cite{leith:74} to account for uncertainty in the initial conditions. In the operational setting, ensemble forecasting of general circulation models (GCMs) has been the principal tool to realize the probabilistic forecasting approach at both the NCEP and ECMWF since early 1990s \citep{kalnay:03}. However, despite advances in computational capacity and observation networks, the GCMs poorly represent the variability of some physical processes, such as tropical convection. A significant challenge in representing this process, as pointed out in \cite{fmk:12}, is the failure of current models to adequately represent the interactions between the large-scale atmospheric circulation and the convective-scale cloud processes. 
Such inadequate modeling, which is generally known as \emph{model error}, is the fundamental barrier to improve medium-range and long-term forecasting skill in NWP.

Many parametric modeling approaches have been proposed and successfully applied to compensate for model error in various contexts \citep[for some examples see][just to name a few]{grabowski:01,HMM,fmk:12,mg:13,kms:04,kms:05,mg:13,mh:13,hmm:14}.  These approaches typically rely on physical insights into specific problems for choosing the appropriate parametric form. In fact, it was rigorously shown in a simple context that it is possible to fully compensate for model error arising from unresolved scales; in the sense that one can simultaneously obtain optimal data assimilation and climatological statistical estimates assuming that one has access to the \emph{right} stochastic parametric model \citep{BH:14}. While stochastic parameterization techniques have been successful for many particular problems, this approach also has another limitation beyond the selection of the parametric form. Namely, determining the parameters in these models from a limited amount of noisy data can be nontrivial, especially when the parameters are not directly observed. This was shown in \cite{BH:14}, which found that even when the appropriate parametric form is known, the choice of the stochastic parameter estimation scheme could have a significant effect on the results.  While some parameter estimation methods have been developed which consistently provide accurate data assimilation and climatological statistics \citep[for some examples see][]{hmm:14,bs:13,ZH14}, these methods are numerically more expensive than the standard linear regression fitting method \citep{amp:13}. 

Recently, a nonparametric method for learning a dynamical system from a training data set was introduced in \cite{BH14UQ} and significantly generalized in \cite{BGH14}. These methods essentially provide a black-box model for forecasting distributions based on the training data. Moreover, it was shown in \cite{BH15PHYSD} that the nonparametric approach of \cite{BGH14} provides high forecasting skill, and is competitive with some physics-based parametric models for predicting energetic modes of turbulence dynamical systems using relatively small training data sets (on the order of $10^3$ data points).  The results of \cite{BH15PHYSD} also demonstrate that, combined with the geometric time-delay embedding theory of \cite{DMDC}, the nonparametric method of \cite{BGH14} is effective even when the training data set is corrupted by noise. Building on these existing results, the nonparametric modeling approach developed in \cite{BGH14} will be the nonparametric core of the \emph{semiparametric} modeling paradigm introduced in this paper. 

The novel aspect of the present paper, beyond \cite{BGH14}, is to overcome the practical limitation of the nonparametric model to low dimensional dynamics.  We overcome this limitation by assuming that we have an approximate or incomplete parametric model, and using nonparametric methods to fill in the missing components, in other words, to correct the model error.  By assuming that the parametric model captures most of the variability, so that the model error is low-dimensional, the nonparametric model becomes feasible.  Simultaneously, from the perspective of the modeling community, the goal of semiparametric modeling is to overcome model error by using historical data to `correct' the existing physical models.  These two perspectives indicate that semiparametric modeling has the potential to seamlessly blend the strengths of the parametric and nonparametric modeling approaches, thereby overcoming their complementary weaknesses.

Semiparametric modeling is well known in the statistics literature as a flexible work-around for overcoming the curse-of-dimensionality \citep{semiparamBook}.  Classical nonparametric models such as histograms or kernel density estimates can reconstruct an unknown \emph{time-independent} density from a training data set, however the amount of data needed grows exponentially in the dimensionality of the data. Alternatively, estimating parameters for a parametric statistical model can often recover a high-dimensional density from a relatively small data set. However, imposing the rigid structure of the parametric form opens up the possibility of model error.  Semiparametric models in statistics attempt to take an intermediate road by choosing a parametric form but allowing the parameters to vary and building nonparametric models for the parameters.  Choosing the structure of the parametric form allows one to encode the prior knowledge about the data set, hopefully reducing the intrinsic dimensionality of the parameter space to the point where a nonparametric model can be used to capture the remaining variability. In this paper, we will extend this statistical idea of semiparametric modeling to estimating \emph{time-dependent} densities of dynamical systems. 

Our plan is to demonstrate that one can use a nonparametric model constructed using the method of \cite{BGH14} instead of using a problem specific parametric model to compensate for the model error.  Combining the partial parametric model with the nonparametric model requires new semiparametric forecasting and filtering algorithms.  However, an important consideration is that we wish to maintain as much of the current parametric ensemble forecasting and filtering framework (as used in NWP) as possible. To achieve these goals, in Section \ref{problem} we will define the form of the model error which we will be able to address here, and we briefly review some simple stochastic parametric models which have been successful for certain types of model error. We will assume that the model error can be described by dynamically varying certain parameters in the parametric model, and that the evolution of these parameters is independent of the state variables of the parametric model. 

In Section \ref{forecast} we introduce a semiparametric forecasting algorithm which combines the nonparametric forecast of \cite{BGH14} with a standard ensemble forecast method for the parametric model. For simplicity and clarity, in Section \ref{forecast} we assume that a training data set of the varying parameters is available and that we are given noisy initial conditions for forecasting. In Sections \ref{findp} and \ref{filter}, we will discuss additional strategies for dealing with the more realistic scenario when we only have noisy observations of the state variables of the parametric model, which is the common situation in practice. In particular, in Section \ref{findp} we use an adaptive filtering method developed in \cite{bs:13,BH:14} to extract a time series of the varying parameters from the noisy observations. In Section \ref{filter} we introduce a semiparametric filter which combines a Ensemble Kalman Filter (EnKF) for the parametric model with the nonparametric model (learned from the time series recovered in Section \ref{findp}) in order to find initial conditions from the noisy observations. As a proof-of-concept example, we will demonstrate the semiparametric filter and forecast on the Lorenz-96 model with model error arising from a parameter which evolves according to low-dimensional deterministic chaotic or stochastic models. We close this paper with a short summary in Section~\ref{summary}.

\section{Problem Statement and Background}\label{problem}

We consider the problem of filtering and forecasting in the presence of model error.  We assume that we are given a noisy time series of observations from a known dynamical model, 
\begin{align}\label{model} \dot x = f(x,\theta), \end{align}
with observation function,
\begin{align}\label{obs} y = h(x,\theta), \end{align}
depending on parameters $\theta$ which evolve according to an unknown stochastic model,
\begin{align}\label{npmodel} \dot \theta = g(\theta,\dot W), \end{align}
where $\dot{W}$ denotes a white noise process. We assume that the evolution of $\theta$ is ergodic on a physical parameter domain $\mathcal{M}$ which may be a subset of the parameter space, $\theta \in \mathcal{M} \subset \mathbb{R}^n$. Moreover, we assume that the distribution of $\theta$ evolves according to a Fokker-Planck equation with a unique equilibrium density $\peq(\theta)$.
A key assumption in this paper is that the evolution of the parameters, $\theta$, does not depend on the state $x$.  This says that the model error is homogeneous with respect to the state space, which is a restrictive assumption for practical applications.  The key difficulty in solving the \emph{conditional} or \emph{heterogeneous} problem is that if $g$ was allowed to vary with $x$ there would typically not be enough data at each state $x$ to fit a nonparametric model; we discuss this further in Section~\ref{summary}.  

With the above problem statement in mind, we will first consider an idealized scenario where we have a training time series of the parameters $\theta$. In this first scenario we will also bypass the filtering step and assume that we are given an initial state $(x(t),\theta(t))$ at some time $t$ from which we wish to forecast the state into the future. Without knowing the model in \eqref{npmodel}, our goal is to first use the training time series to learn an approximate model $\tilde g \approx g$ for $\theta$. Subsequently, we combine this approximate model $\tilde g$ for $\theta$ with the parametric model \eqref{model} in order to forecast the initial condition $(x(t),\theta(t))$.  In the remainder of this section we will review several standard methods for building approximate models, $\tilde g$.

Given a discrete time series of historical values of the parameters $\{\theta_i = \theta(t_i)\}_{i=1}^N$ there are several standard approaches for building a forecast model for $(x,\theta)$. The simplest method is to simply ignore the training data, and hold the value of $\theta$ constant at the initial parameter value $\theta(t)$, so that $\tilde g = 0$, and we will refer to this method as the \emph{persistence model}.  The persistence model integrates the system \eqref{model} with $\theta(t+\tau)=\theta(t)$ held constant at the initial value $\theta(t)$, in order to obtain the forecast of $x$.  When the evolution of $\theta$ is deterministic and very slow relative to $x$, the persistence model can give reasonable short term forecasts, assuming that the system is still stable with $\theta$ fixed.  Of course, if the evolution of $\theta$ is very fast, the persistence model will quickly diverge from the true state.  Moreover, the persistence model will typically not capture the correct long term mean of the true system, leading to a very biased long term forecast which can easily be worse than simply predicting the mean value of $x$.  

When the evolution of $\theta$ is very fast, we can assume that, relative to $x$, the parameters $\theta$ quickly equilibrate to their climatological distribution.  Assuming that the evolution of $\theta$ is ergodic and that we have a sufficiently long training data set $\{\theta_i\}$, we can emulate the equilibrium density by simply drawing random parameters from the historical data.  To use this in the forecast, for each integration step of the forecast model \eqref{model} we draw a random value of $\theta$ from the historical data; this strategy is similar to the \emph{heterogeneous multiscale method (HMM)} framework \citep{HMM}. For the long term forecast, the HMM gives an unbiased forecast model which recovers the correct equilibrium density, however for the short term forecast, the HMM completely ignores the known initial parameter value $\theta(t)$.  

So far, the persistence model assumes the evolution of the parameters is slow and ignores the training data, whereas the HMM assumes the evolution of the parameters is fast and ignores the initial state $\theta(t)$.  It turns out that blending these two approaches is quite challenging.  A method which attempts to blend these, in the sense that it moves the initial state $\theta(t)$ toward the long term mean $\hat \theta = \frac{1}{N}\sum_{i=1}^N \theta_i$ at an appropriate speed, was introduced as part of the MTV strategy for reduced climate modeling \citep{mtv:99}. In \cite{mtv:99} they approximate the interaction between the unresolved scales with a parametric model involving a linear damping term and white noise stochastic forcing. In our context, since $g$ is assumed to be independent of $x$, we simply have a linear stochastic model,
\begin{align}\label{msm} \dot \theta = \tilde g(\theta,\dot W) = \alpha (\hat \theta - \theta) + \sigma \dot W, \end{align}
where $\alpha = 1/T_\theta$ is determined by the correlation time $T_\theta$ of the historical data and the stochastic forcing $\sigma = \sqrt{2\alpha \textup{var}(\theta)}$ is determined by the variance of the historical data. This approach is known as the \emph{mean stochastic model (MSM)} \citep[see][]{mgy:10}. Assuming that the initial density for the parameters is Gaussian with mean given by the initial state $\theta(t)$ and an initial variance $s(t)$ (possibly coming from a Kalman filter), the density of the MSM forecast at time $t+\tau$ is a Gaussian and its mean and variance can be determined explicitly. Using the analytic solution for the linear (and Gaussian) model in \eqref{msm}, we can generate random samples from the Gaussian forecast at each integration step of the ensemble forecast.  While the MSM effectively uses the initial state, it only captures two moments (namely the mean and covariance) and the two-time correlation statistics of the climatological distribution. 

In our numerical experiments below, we will compare the forecasting skill of the semiparametric model with the three approaches mentioned above, namely the persistence, HMM, and MSM models, in addition to the perfect model. There are of course other more sophisticated parametric stochastic models, and we should point out that a comparison with these methods was already made in the context of nonparametric forecasting of turbulent modes in \cite{BH15PHYSD}.  In this paper we assume that all of the physical knowledge has already been incorporated into the parametric model, so that we have absolutely no knowledge of the system \eqref{npmodel} which could be used to suggest a better parametric form.

\section{Semiparametric Forecasting}\label{forecast}

We will assume that the model for $x$ is nonlinear and potentially high-dimensional and chaotic so that the most practical method of forecasting $x$ is simply to integrate an ensemble of initial states forward in time using the model \eqref{model}. Semiparametric forecasting consists of using the training data to learn a nonparametric model for $\theta$ and combining this model with the dynamical model \eqref{model} to improve the forecast of an initial state $(x(t),\theta(t))$. 

To simplify the discussion in this section, we assume that we have a training data set consisting of a historical time series $\{\theta_i=\theta(t_i)\}$ of the parameters and we are given an initial state $x(t)$ and the initial parameter values $\theta(t)$ at some time $t$, from which we would like to forecast the state $x$ into the future. We will consider the more realistic situation, where the training data set and initial conditions for $\theta$ is not available, in Sections \ref{findp} and \ref{filter}. In the remainder of this section, under the current idealized scenario, we introduce the semiparametric forecasting approach.  In Section \ref{nonparametricforecast} we briefly review the nonparametric model of \citep{BGH14} for $\theta$. In Section \ref{combinemodels} we show how to combine this nonparametric model with the known parametric model, $f$, to perform a semiparametric forecast.  Finally, in Section \ref{perfectexample} we demonstrate the semiparametric forecast on an example based on a Lorenz-96 model, where we introduce a complex model error which is governed by the Lorenz-63 model.

\subsection{Step 1: Forecasting parameters with the nonparametric model}\label{nonparametricforecast}

We assume that the true dynamics of the parameters $\theta(t)$ are constrained to a physical parameter domain $\mathcal{M}\subset\mathbb{R}^n$, and evolve according to a stochastically forced dynamical system \eqref{npmodel}.  Notice that we allow the data to be constrained to a domain $\mathcal{M}$ inside the parameter space $\mathbb{R}^n$, so even when the dimension, $n$, of the parameter space is high, the data requirements will only depend on the dimension of $\mathcal{M}$. We emphasize that the domain $\mathcal{M}$, as well as $g$, are unknown and will be learned implicitly from the training data set. Moreover, the parameters $\theta$ may include variables which do not appear explicitly in the model \eqref{model}, but we assume that our training data set only consists of observable variables which appear explicitly in \eqref{model}. In order to compensate for unobserved variables in the dynamics of the parameters $\theta$ which appear in \eqref{model}, we will apply a time-delay embedding to the available variables \citep{embedology,stark1,stark2,DMDC,giannakisMajda,GiannakisPNAS,BH15PHYSD}.  

Given a time series $\theta_i = \theta(t_i)$ sampled at discrete times $\{t_i\}_{i=1}^{N}$ we are interested in constructing a forecasting model so that given an initial density $p(\theta,t)$ at time $t$ we can estimate the density $p(\theta,t+\tau)$ at time $t+\tau$, where $\tau>0$.  We assume that $\theta_i$ is already the delay coordinate embedding of the available parameters which appear explicitly in the parametric model \eqref{model}.  Probabilistically, the evolution of the density, $p$, in the forecasting problem is governed by the Fokker-Planck equation,
\begin{align}
\frac{\partial p}{\partial t} = \mathcal{L}^*p, \quad p(\theta,t) = p_t(\theta),\label{FokkerPlanck}
\end{align}
where $\mathcal{L}^*$ is the Fokker-Planck operator, which we assume exists for the system \eqref{npmodel}. We note that the equilibrium distribution, $\peq(\theta)$, of the underlying dynamics \eqref{npmodel} satisfies, $\mathcal{L}^*\peq = 0$.

The nonparametric forecasting algorithm introduced in \cite{BGH14} is called the \emph{diffusion forecast}.  The central idea of the diffusion forecast is to project the forecasting problem \eqref{FokkerPlanck} onto a basis of smooth real-valued functions $\{\varphi_j(\theta)\}$ defined on the parameter domain $\mathcal{M}$. We will discuss below how the basis $\{\varphi_j\}$ will be estimated using the training data; for now we assume that the basis functions are known.  In order to find the forecast density $p(\theta,t+\tau)$ from the initial density $p(\theta,t)$, we first project this function onto the basis $\{\varphi_j\}$ by computing the projection coefficients $c_j(t) = \langle p(\cdot,t),\varphi_j\rangle$. \comment{where the inner product $\langle \cdot,\cdot\rangle$ will be numerically approximated by a Monte-Carlo integral.} We can then forecast the coefficients $\vec c(t)$ with a linear map $\vec c(t+\tau) = A\vec c(t)$, whose components are defined by $A_{lj}  \equiv \mathbb{E}\left[\langle \varphi_j, S\varphi_l\rangle_{\peq}\right]$, where $S$ is the discrete \emph{shift map} defined by $S\varphi_l(\theta_i) = \varphi_l(\theta_{i+1})$. \comment{Note that the expectation here is with respect to realizations of \eqref{SDE}.}It was shown in \citep{BGH14} that the linear map $A$ is exactly the forecasting operator (which takes the density $p(\theta,t)$ to the forecast $p(\theta,t+\tau)$) represented in the basis $\{\varphi_j\}$. The time step, $\tau$, of the forecasting operator is determined by the sampling time $\tau \equiv t_{i+1}-t_i$ of the training data.  Finally, we can reconstruct the forecast density by,
\begin{align}
p(\theta,t+\tau) = \sum_j c_j(t+\tau)\peq(\theta)\varphi_j(\theta).\nonumber
\end{align} 
These formula were all derived in \citep{BGH14}, and we now summarize the nonparametric forecasting approach with the following diagram, 
\[ \arraycolsep=1.4pt\begin{array}{clc}
\hspace{-15pt}p(\theta,t) \hspace{0pt}&\xdashrightarrow{\ \ \ \ \textup{Nonparametric Forecast} \ \ \ \ }&\hspace{-15pt} p(\theta,t+\tau) \\ \\
\hspace{5pt}\left\downarrow\rule{0cm}{.5cm}\right.   \scriptstyle{\left<p,\varphi_j\right>}&&\hspace{10pt}\left\uparrow\rule{0cm}{.5cm}\right. \scriptstyle{\sum_j c_j \varphi_j\peq}\\ \\
\hspace{-15pt}\vec c(t) \hspace{0pt}&\xrightarrow{\ \ \ A_{lj} \equiv \mathbb{E}[\langle \varphi_j, S\varphi_l\rangle_{\peq}] \ \ \ }& \vec c(t+\tau) = A \vec c(t).
\end{array} \] 
In the above diagram, the nonparametric forecasting algorithm follows the solid arrows: projecting onto the basis, applying the linear map $A$ (which computes the forecast), and then reconstructing the forecast density. 

Given a basis $\{\varphi_j\}$ and an estimate of the equilibrium density $\peq$, it was shown in \citep{BGH14} that the above inner products can be estimated as Monte-Carlo integrals on the training data set $\{\theta_i\}$ given by,
\begin{align}
c_j(t) &\equiv \left<p(\cdot,t),\varphi_j\right>  \approx \frac{1}{N}\sum_{i=1}^N p(\theta_i,t)\varphi_j(\theta_i)/\peq(\theta_i)  \label{mc1}\\
\hat A_{lj} &\equiv \left<\varphi_j, S \varphi_l \right>_{\peq}\approx \frac{1}{N-1}\sum_{i=1}^{N-1} \varphi_j(\theta_i)\varphi_l(\theta_{i+1}).\label{mc2}
\end{align}
Notice that we only need estimates of the basis $\{\varphi_j\}$ and the density $\peq$ evaluated on the training data set.
Of course, in practice we will truncate the projections onto a finite number $M$ of the basis functions, $\varphi_j$, so that we implicitly project the forecast density into the span of the first $M$ basis functions, $\{\varphi_j\}_{j=1}^M$. Notice that by projecting and reconstructing the density $p$ in the basis $\{\varphi_j\}$ (see the diagram above), we have projected the density in the ambient space onto the physical domain $\mathcal{M}$, as determined by the historical data set.  In fact, $\{\varphi_j\}$ is a generalized Fourier basis and since we only use $M$ basis elements, it is possible to have Gibbs phenomenon oscillations in the reconstructed density, which can lead to negative values.  In order to avoid these negative values, whenever we reconstruct the density we will always take the max of the reconstructed values and zero, and then renormalize the density by dividing by,
\begin{align} \label{normalizationfactor} Z = \int_{\theta \in \mathcal{M}} p(\theta,t) \approx \frac{1}{N} \sum_{i=1}^N p(\theta_i,t)/\peq(\theta_i) \end{align}
which estimates the normalization factor.

The final piece of the nonparametric forecasting algorithm is choosing an appropriate basis $\{\varphi_j\}$. It was shown in \cite{BGH14} that the optimal basis for minimizing the error between the estimate $\hat A$ of the matrix $A$ are the eigenfunctions of an elliptic operator $\hat{\mathcal{L}}$.  The operator $\hat{\mathcal{L}}$ is the generator of a stochastic gradient flow in the potential field $U(\theta) = -\log \peq(\theta)$, where $\peq$ is the equilibrium density of \eqref{npmodel}. By choosing this basis of eigenfunctions, it was shown in \cite{BGH14} that the difference between the estimate $\hat A_{lj}$ and the true coefficient $A_{lj}$ is an error of order $\tau/N$ in probability. Most importantly, these eigenfunctions, $\{\varphi_j\}$, and the density $\peq$, can be estimated with the diffusion maps algorithm \citep{diffusion,BH14VB}.

Intuitively, the diffusion maps algorithm uses the data set $\{\theta_i\}_{i=1}^N$, to approximate the operator $\hat{\mathcal{L}}$.  For any function $F:\mathcal{M}\to\mathbb{R}$ we can represent $F$ on the data set $\{\theta_i\}$ by an $N\times 1$ vector $\vec F_i = F(\theta_i)$. The operator $\hat{\mathcal{L}}$ is then approximated by an $N\times N$ matrix $L$, in the sense that the matrix-vector product $L\vec F$ has entries $(L\vec F)_i \approx \hat{\mathcal{L}}(F)(\theta_i)$ so that the vector $L\vec F$ approximates the function $\hat{\mathcal{L}}F$ on the data set.  The algorithm for constructing the matrix $L$, and the theory proving its convergence to $\hat{\mathcal{L}}$ in the limit of large data, were established in \cite{BH14VB} for sparsely sampled data; generalizing the seminal result of \cite{diffusion}. The eigenvectors of the matrix $L$ approximate the desired eigenfunctions evaluated on the training data set, as required in \eqref{mc1} and \eqref{mc2}. The diffusion maps algorithm simultaneously provides a kernel density estimate of $\peq$ evaluated on the training data set as required in \eqref{mc1} and \eqref{normalizationfactor}. As a result, the reconstructed forecast density, $p(\theta,t+\tau)$, is represented by its values on the training data set, $p(\theta_i,t+\tau)$.  In this article, we use the algorithm of \cite{BH14VB} since the equilibrium density, $\peq$, may take values arbitrarily close to zero, which would imply regions of sparse sampling on $\mathcal{M}$. For a brief overview of the theory of \cite{BH14VB}, see the Supplementary Material of \cite{BGH14}, and for a simple guide to the algorithm see the appendix of \cite{BH15PHYSD}.

\subsection{Step 2: Combining the nonparametric and parametric forecast models}\label{combinemodels}

In this section, we introduce the semiparametric forecasting scheme which combines the parametric model \eqref{model} for $x$ and the nonparametric model of Section \ref{nonparametricforecast} for $\theta$. We assume that we are given an ensemble of $K$ initial conditions $\{x^k(t)\}_{k=1}^K$ which represent the distribution of the variable $x$ at the initial time $t$, and an initial density $p(\theta,t)$ for the parameters, and our goal is to forecast these two quantities to the future time $t+\tau$.

We first describe how to forecast the ensemble $\{x^k\}$. Since the model for $x$ is parametric of the form $f(x,\theta)$, the evolution of $x$ depends on the parameters $\theta$.  In order to forecast each ensemble member, $x^k$, with the parametric model, we need to produce samples $\{\theta^k\}_{k=0}^K$ of the density $p(\theta,t)$. We will sample the density $p(\theta,t)$ using the standard rejection method, where the reference density is the equilibrium density $\peq$, which is also the sampling measure by our ergodicity assumption on $\theta$. To perform the rejection sampling, we set $P = \max_i\{p(\theta_i,t)/\peq(\theta_i) \}$ to be the upper bound of the ratio between the two distributions. Notice that since both densities are known on the historical training data set $\{\theta_i\}$ we can easily compute this optimal upper bound.  Since the historical training data points $\{\theta_i\}$ are samples of $\peq$, we can easily produce a random sample of the density $\peq$ by randomly choosing a training data point $\theta_m$ where $m \in \{1,...,N\}$ is a uniformly random index.  We then choose a uniform random number $\xi \in [0, 1]$ and we accept $\theta_m$ as a sample of $p(\theta,t)$ if $\xi < p(\theta_m,t)/(P\peq(\theta_m))$ and otherwise we reject $\theta_m$ and redraw until we accept.  We repeat this rejection sampling until we have $K$ samples $\{\theta^k(t)\}_{k=1}^K$ from the density $p(\theta,t)$.  Each ensemble member $\{x^k(t)\}$ is then paired with the corresponding sample from $p(\theta,t)$ to form the combined ensemble $\{(x^k(t),\theta^k(t))\}$.  We can now integrate the parametric model $\dot x = f(x,\theta)$, using each ensemble state $(x^k(t),\theta^k(t))$, for time $\tau$ in order to find the forecast ensemble $\{x^k(t+\tau)\}$ at time $t+\tau$.  

The second part of the semiparametric forecasting algorithm is exactly the diffusion forecasting scheme described in \eqref{nonparametricforecast}. To repeat, we project $p(\theta,t)$ onto the basis $\{\varphi_j\}$ to find the coefficients $c_j(t)$ as in \eqref{mc1}. We then forecast the density on the parameters using the matrix $\hat A$  from \eqref{mc2} so that $c(t+\tau) = \hat Ac(t)$ and we can then reconstruct the density at time $t+\tau$. Subsequently, we sample $p(\theta,t+\tau)$ using the same rejection sampling procedure to obtain an ensemble $\{\theta^k(t+\tau)\}$ which we then pair with $\{x^k(t+\tau)\}$ to form $\{(x^k(t+\tau),\theta^k(t+\tau))\}$ and again integrate to find $\{x^k(t+2\tau)\}$. We can then repeat the procedure by finding $c(t+2\tau) = Ac(t+\tau)=A^2c(t)$, reconstructing the density $p(\theta,t+2\tau)$ and so on for as long as we would like to forecast. Therefore, in addition to the nonparametric forecasting described before, the \emph{semiparametric forecasting algorithm} consists of sampling the nonparametric forecast density at each forecast step and using these samples for integrating an ensemble forecast of the parametric model.

We summarize the semiparametric forecasting algorithm with the following diagram,
\[ \arraycolsep=1.4pt\begin{array}{clc}
(x^k(t),\theta^k(t))\hspace{0pt}&\xrightarrow{\ \ \ \ \ \ \ \ \ \ \dot x = f(x,\theta) \ \ \ \ \ \ \ \ \ \ }&(x^k(t+\tau),\theta^k(t+\tau))\\ \\
\hspace{0pt}\left\uparrow\rule{0cm}{.5cm}\right.   \scriptstyle{\theta^k(t)}&&\hspace{0pt}\left\uparrow\rule{0cm}{.5cm}\right. \scriptstyle{\theta^k(t+\tau) }\\ \\
\hspace{-15pt}p(\theta,t) \hspace{0pt}&\xdashrightarrow{\ \ \ \ \textup{Nonparametric Forecast} \ \ \ \ }&\hspace{-15pt} p(\theta,t+\tau) \\ \\
\hspace{5pt}\left\downarrow\rule{0cm}{.5cm}\right.   \scriptstyle{\left<p,\varphi_j\right>}&&\hspace{10pt}\left\uparrow\rule{0cm}{.5cm}\right. \scriptstyle{\sum_j c_j \varphi_j\peq}\\ \\
\hspace{-15pt}\vec c(t) \hspace{0pt}&\xrightarrow{\ \ \ A_{lj} \equiv \mathbb{E}[\langle \varphi_j, S\varphi_l\rangle_{\peq}] \ \ \ }& \vec c(t+\tau) = A \vec c(t).
\end{array} \]
Notice that the nonparametric model iterates independently from the parametric model, which follows from the assumption that the model \eqref{npmodel} for $\theta$ is independent of the state variable $x$.  Furthermore, the dashed line is not taken directly but is a summary of the path of solid lines which uses the projection onto the basis $\{\varphi_j\}$.  The parametric model is used to forecast the ensemble of states $x^k(t)$ and at each step the ensemble of parameters $\theta^k(t)$ are updated by sampling the density $p(\theta,t)$ from the nonparametric forecast. As in Section \ref{nonparametricforecast} the initial density $p(\theta,t)$, which can be arbitrary, will be evaluated on the data set, and all subsequent densities $p(\theta,t+m\tau)$ are represented by the values $p(\theta_i,t+m\tau)$.

We should note that the ensemble members $\theta^k(t+m\tau)$ are not actually time series solutions of \eqref{npmodel}, since at each forecast step we choose a completely new set of random samples independently from the samples of the previous step.  Producing a consistent time series for each ensemble forecast $\theta^k(t+m\tau)$ would require sampling the conditional density at each time given the sample at the previous time. This would allow more complex integration schemes to be considered for combining the samples $\{\theta^k\}$ with the ensemble $\{x^k\}$, however, for a small ensemble size it would result in a more biased forecast.

\subsection{Application of semiparametric modeling to a chaotic system with model error}\label{perfectexample}

\begin{figure*}
\centering
\includegraphics[width=0.98\textwidth]{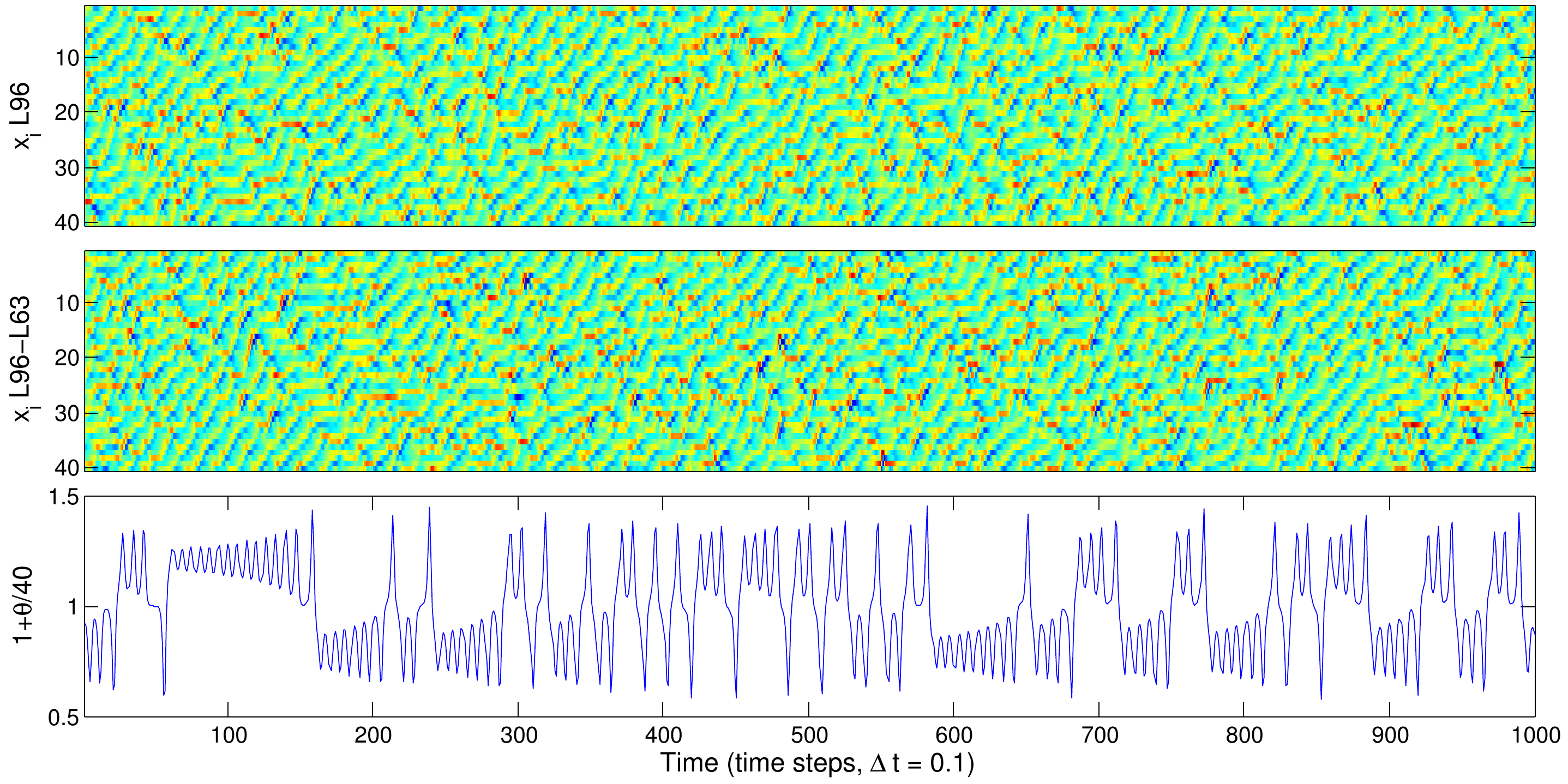}
\caption{\label{L96L63comp} Comparison of the Lorenz-96 (top) model dynamics with the modified system \eqref{l96l63} (middle) and the coefficient $(\theta/40+1)$ (bottom).}
\end{figure*}

In this example we will introduce model error into the 40-dimensional Lorenz-96 \citep{lorenz:96} system,
\begin{align}\label{l96}
\dot x_i &= x_{i-1}x_{i+1}-x_{i-1}x_{i-2} - x_i + 8, 
\end{align}
where the indices are taken modulo 40, which is known to have chaotic dynamics.  Notice that the system is dissipative in the sense that the evolution of the energy-like quantity, $S \equiv \sum_i x_i^2/40$ is simply $\dot S = -2S+2/5 \sum_i x_i$.  We will now introduce a model error by assuming that the evolution of the true system is governed by,
\begin{align}\label{l96l63}
\dot x_i &= f(x_i,\theta) = \theta x_{i-1}x_{i+1}-x_{i-1}x_{i-2} - x_i + 8 \nonumber \\ 
\theta &= (a_1/40+1) \nonumber \\
\dot a_1 &=10(a_2 - a_1)/\epsilon,  \\
\dot a_2 &=(28 a_1 - a_2 - a_1 a_3)/\epsilon,\nonumber \\
\dot a_3 &=(a_1 a_2 - 8a_3/3)/\epsilon, \nonumber
\end{align}
which we call the L96-L63 system, where we replace the constant coefficient in the first quadratic term of \eqref{l96} with a parameter $\theta$. The parameter $\theta$ is an appropriate rescaling of the first component, $a_1$, of the chaotic three-dimensional Lorenz-63 \citep{L63} system. The system \eqref{l96l63} does not have the dissipative property of \eqref{l96} but is empirically stable in long numerical simulations. Empirically, we found that the system \eqref{l96l63} became very unstable if $\theta$ was allowed outside the interval $[0.5, 1.5]$. The constant $\epsilon$ is included to allow differences in the time-scale between the parametric model for $x$ and the parameters $\theta$.  In this section we will use $\epsilon=1$ however we will consider faster ($\epsilon<1$) and slower ($\epsilon>1$) time-scales in subsequent sections.  In Figure \ref{L96L63comp} we compare the dynamics of these two systems. Notice that the spatio-temporal patterns of these two very different dynamical systems appear deceptively similar.  

While this example is a proof of concept for semiparametric forecasting where we assume we are given a training time series, in real applications we can at most extract the time series of $\theta$ which appears explicitly in the parametric model.  In Section~\ref{findp} we describe a method of extracting a time series of $\theta$ from noisy observations, and we will see that we cannot hope to recover the three hidden variables $\{a_1, a_2, a_3\}$. Since the hidden variables do not appear explicitly in the parametric model $f$ in \eqref{l96l63}, we will train the nonparametric model using only a time series of 5000 values of $\theta_i = \theta(t_i)$ with spacing $\Delta t = t_{i+1}-t_i = 0.1$.  In order to account for the hidden variables in the nonparametric model, we perform a time delay embedding using 4-lags, meaning that we replace the one-dimensional time series $\theta_i$ with the time-delay embedding coordinates, $(\theta_i,\theta_{i-1},...,\theta_{i-4})^\top$, prior to building the nonparametric model.

In order to compare the forecasting skill of various methods with imperfect initial conditions, we introduce a Gaussian perturbation to the true state,
\[ (\tilde x(t),\tilde \theta(t)) = (x(t),\theta(t)) + \eta_t, \hspace{10pt} \eta_t \sim \mathcal{N}(0,C(t)), \]
where the perturbation $\eta_t$ has mean zero and variance equal to 0.1\% of the long term variance of each variable.  In other words, we define the covariance matrix,
\begin{align}\label{covmat} C(t) = \left(\begin{array}{cc} C_{xx}(t) & C_{x\theta}(t) \\ C_{\theta x}(t) & C_{\theta \theta}(t) \end{array}\right), \end{align}
where $C_{xx}(t)$ is diagonal matrix of size $40\times 40$ and $C_{\theta\theta}(t)$ is $1\times 1$ for this example; and the diagonal entries of these sub-matrices are equal to 0.1\% of the variance of the respective variables in \eqref{l96l63}. In this numerical experiment, we assume that the cross covariances $C_{x\theta}(t)=C_{\theta x}(t) = \vec 0$, and moreover, notice that the covariance matrix $C(t)$ is actually the same for every initial time, $t$, in this experiment. In Section~\ref{filter}, the semiparametric filter solutions will produce time varying covariances and cross-covariance which are nonzero. Each forecast method will be given the perturbed initial state $(\tilde x(t),\tilde \theta(t))$ and the covariance matrix $C(t)$, and each method will attempt to forecast the true state $x(t+m\tau)$ for $m=0,1,...,50$ time steps.

Typically we are interested in problems where only a small number $K$ of ensemble members, $\{x^k(t)\}_{k=1}^K$, can be integrated; and a good choice are the sigma points of the covariance ellipse \citep{julier2004unscented}.  The sigma points are given by the adding and subtracting the columns of the symmetric positive definite square root of the covariance matrix $C_{xx}(t)$, to the initial state estimate $\tilde x(t)$. For the persistence model, we simply fix $\theta = \tilde \theta(t)$ and apply the ensemble forecast to the parametric model with this fixed value of $\theta$. For the MSM parameterization, we used the analytic solutions to the Ornstein-Uhlenbeck model (fit to the training data $\{\theta_i\}$) to forecast the initial mean $\tilde \theta(t)$ and covariance $C_{\theta\theta}(t)$ and we used the forecast mean at each discrete time step as the fixed parameter $\theta$ for the ensemble forecast of the parametric model. We also tried using the MSM forecast of the covariance to sample an ensemble of parameters, $\{\theta^k\}$, from the Gaussian forecast density, however we found that this was unstable for this problem.  For the HMM parameterization, we immediately ignore $\tilde\theta(t)$ by drawing samples of initial conditions $\{\theta^k(t)\}_{k=1}^K$ from the equilibrium density, which means that each ensemble member $\theta^k$ is chosen by selecting a random index $m \in \{1,...,n\}$  and setting $\theta^k = \theta_m$ from the training data set. The ensemble of $\{\theta^k\}$ is combined with the ensemble $\{x^k\}$ and the parametric model is used to forecast one step and then the parameter ensemble $\{\theta^k\}$ is resampled and the process is repeated. For the perfect model, we generate an ensemble of $\{(x^k(t), a_1^k(t), a_2^k(t), a_3^k(t)\}_{k=1}^K$ using the sigma points of a $43\times 43$ diagonal covariance matrix with diagonal entries equal to 0.1\% of the long term variance of each variable and then perform the ensemble forecast using the full model \eqref{l96l63}. Note that the initial state for the full model is given by $\tilde a_1(t) = 40(\tilde\theta(t)-1)$ and $(\tilde a_2(t),\tilde a_3(t)) = (a_2(t),a_3(t)) + (\eta_2(t),\eta_3(t))$ where $\eta_2,\eta_3$ are Gaussian with mean zero and variance equal to 0.1\% of the variance of the respective variables.  To initiate the nonparametric forecast in the semiparametric forecasting algorithm, we form an initial density for the parameters using a Gaussian $p(\theta_i,t) = \exp\left(-||\tilde\theta(t)-\theta_i||^2_{C_{\theta\theta}} \right)$ which we normalize by dividing $p(\theta_i,t)$ by the normalization factor \eqref{normalizationfactor}.
We then use the rejection method to sample $\{\theta^k(t)\}_{k=1}^K$ and perform the semiparametric forecast iteratively as described in Section \ref{combinemodels}. 

\begin{figure}[htbp]
\centering
\includegraphics[width=0.45\textwidth]{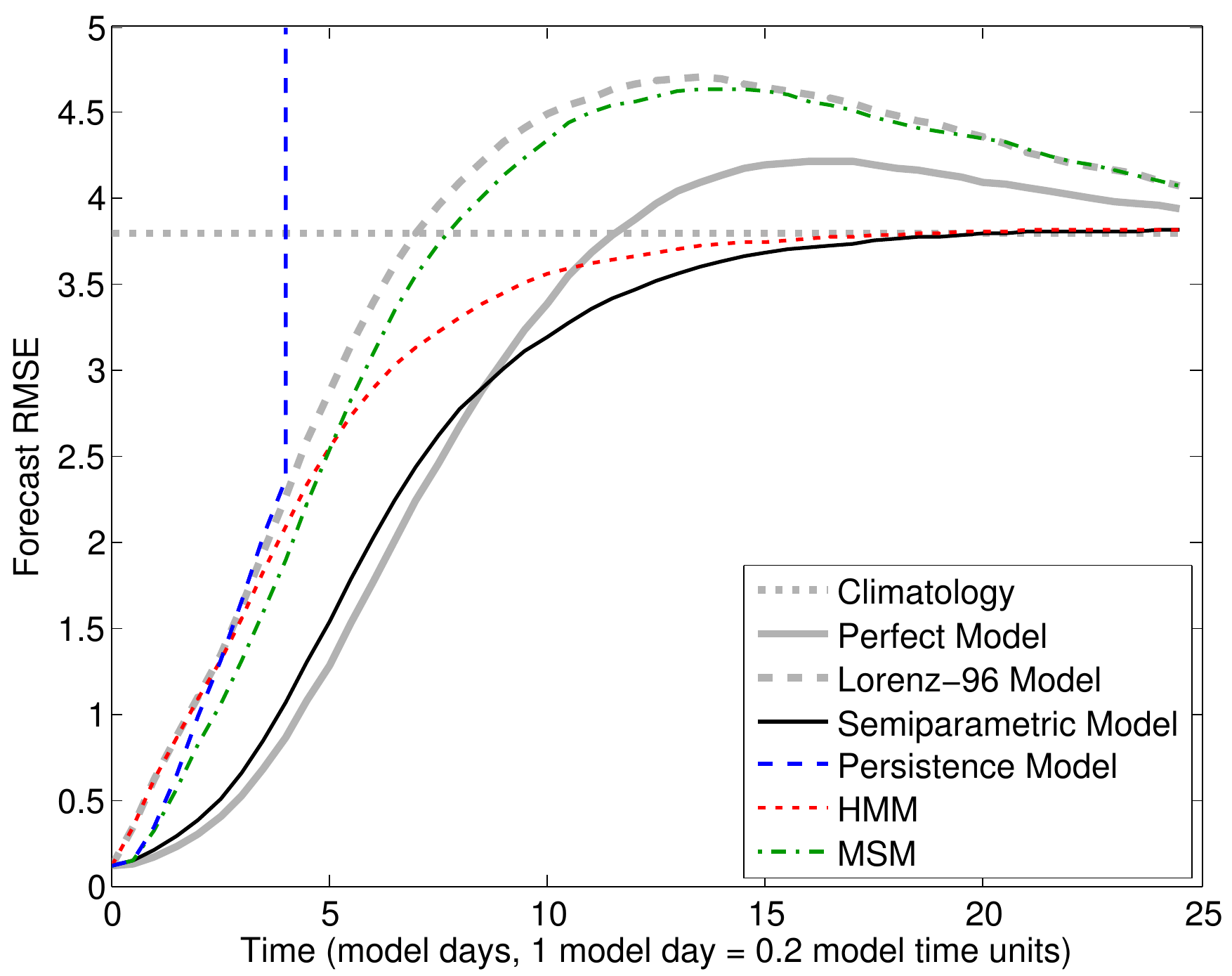}
\includegraphics[width=0.45\textwidth]{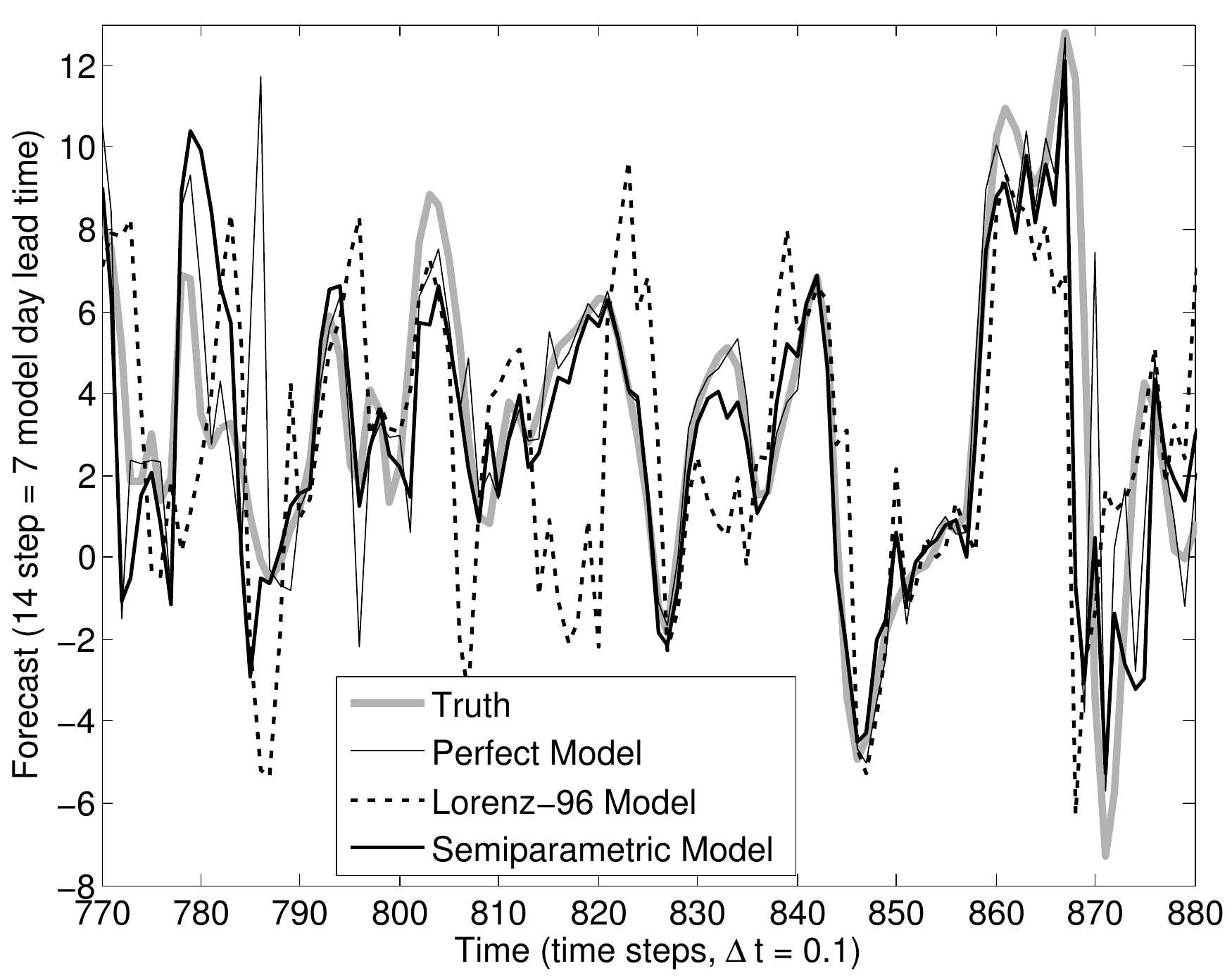}
\caption{\label{L96L63Ex} Comparison of forecasting methods for the model \eqref{l96l63} with $\epsilon = 1$, given perfect training data and perturbed initial conditions for the parameter, $\theta$. The top panel shows the forecast RMS errors as functions of forecast lead time and the bottom panel shows the 7 day lead-time forecast during the verification period of 770-880 time steps.}
\end{figure}

The above methods are applied to $1000$ initial conditions, forecasting for a total of 50 time steps of length $\tau = 0.1$, which corresponds to 5 model time units, or equivalently 25 model days (using the convention of 1 model day being 0.2 model time units of \citep{L63}). The results are shown in Figure \ref{L96L63Ex} with the root mean squared error (RMSE) averaged over the 1000 initial conditions.  With only 86 ensemble members, the ensemble forecast using the perfect model \eqref{l96l63} produces the best short term prediction, but it also seems to produce a biased forecast.  This bias is shown by the RMS error rising above that of the climatological error in the intermediate term beyond 12 model days (which corresponds to $24$ forecast steps, or $2.4$ model time units). In contrast, the RMS error of the semiparametric forecast grows slightly more quickly than the perfect model initially, but the forecast is unbiased in the intermediate term, approaching the climatology without exceeding the climatological error.  In fact, after 8 model days the semiparametric forecast is producing a better forecast than the perfect model.  This is due to the samples of the nonparametric forecast being independent samples of the forecast density, rather than being constrained to follow the paths of the initial ensemble members.  Notice that the HMM, which immediately samples the equilibrium density $\peq$ by drawing random samples from the training data, initially matches the forecasting skill of the unmodified Lorenz-96 model of \eqref{l96}, and in the long term the HMM is unbiased whereas the unmodified model is very biased.  Finally, the persistence model and the MSM use the initial condition $\theta(t)$ to improve the very short term forecast, but in the long term these methods do not capture the true climatology of $\theta$, leading to a biased intermediate to long term forecast.  In fact, the persistence model is so biased that the forecast model diverges catastrophically; this is because the model in \eqref{l96l63} does not conserve energy when $\theta$ is held fixed at a value other than one.  By varying $\theta$ in time, with a mean value of $1$, all the models except for the persistence model are able to give a stable forecast.  In the short term, the semiparametric forecast is able to provide an additional 1-2 model days of equivalent forecast skill compared to the standard methods considered here, and almost matches the perfect model forecast (see the 7 day lead-time forecast in the bottom panel of Figure~\ref{L96L63Ex} during the verification period 770-880 time steps). In the intermediate term forecast, the lack of bias in the semiparametric forecast provides a significant improvement in forecast skill, even compared to the perfect model for this small ensemble size.

\section{Extracting a time series of hidden parameters}\label{findp}

In this section, we will consider the more practical situation in which the historical time series of the parameters $\{\theta_i\}$ is not available. We will show that it is possible to extract such a training data set from noisy observations $y^o = y +\eta =  h(x) +\eta$ where $\eta \sim \mathcal{N}(0,R)$ is a Gaussian observation noise.  Of course, this recovered data set will not be perfect and it will not exactly obey the dynamics of $\theta$ in \eqref{l96l63}. In Section \ref{filter}, we will show that the recovered data set still allows for training a useful nonparametric model which improves the forecast.  If $\theta$ were fixed, or we had a parametric form for the dynamics of $\theta$, then extracting the time series of $\theta$ from the noisy observations $y^o$ would be a standard inverse problem.  However, we would like to recover a time series of values of $\theta$ without knowing anything about the evolution of $\theta$.  This type of inverse problem should require at a minimum the same observability conditions as that of finding a fixed parameter $\theta$.  Our goal here is not to give a formal exposition of this inverse problem, but rather to present a proof-of-concept approach which shows that extracting a time series of $\theta$ is possible in a reasonably difficult context, and that the imperfectly recovered time series of $\theta$ can still be used to improve forecasting via a semi-parametric model.

The approach we take here follows closely the standard state augmentation approach for parameter estimation.  We will recover the time series of $\theta$ as part of an ensemble Kalman filter applied to recover the state variable $x$ from the noisy observations $y$. This approach requires augmenting \eqref{model} with a model for $\theta$. If $\theta$ were simply a constant parameter we could use the persistence model, $\dot\theta=0$, and this corresponds to the standard \emph{state augmentation} approach to parameter estimation \citep{friedland:69}.  Instead, because $\theta$ is changing in time, we will assume a white noise model, $\dot\theta = \sqrt{Q_{\theta\theta}} \dot W$.  A typical approach is to empirically tune the value of $Q_{\theta\theta}$ \citep{friedland:82} possibly using some a priori knowledge of the variable $\theta$. Intuitively, the true signal $\theta$ is being treated as a realization of a white noise process, and so a good choice for $Q_{\theta\theta}$ would be the variance of the time series $\{\theta_i\}$.  Of course, since recovering the time series $\theta_i$ is our goal, we do not assume that the variance is known, and instead we use a method of adaptive filtering which can automatically recover $Q_{\theta\theta}$.  The method used here to find $Q_{\theta\theta}$ is detailed in Appendix \ref{QR} and is a version of the approach introduced in \citep{bs:13} and applied in \citep{BH:14}, which is a nonlinear generalization of an earlier approach in \citep{mehra:72}.  We note that a closely related technique was simultaneously developed in \citep{hmm:14} and is refined in \citep{ZH14}.  Intuitively, these methods use the error between the predicted and actual observations (which is also known as the innovation), to determine the amount of noise which must have been added to the state in order to explain this error.  By analyzing correlations between the innovations at different lags it is possible to find the observation covariance $R$ as well as observable parameters of the dynamical noise, such as $Q_{\theta\theta}$.

\begin{figure}[htbp]
\centering
\includegraphics[width=0.45\textwidth]{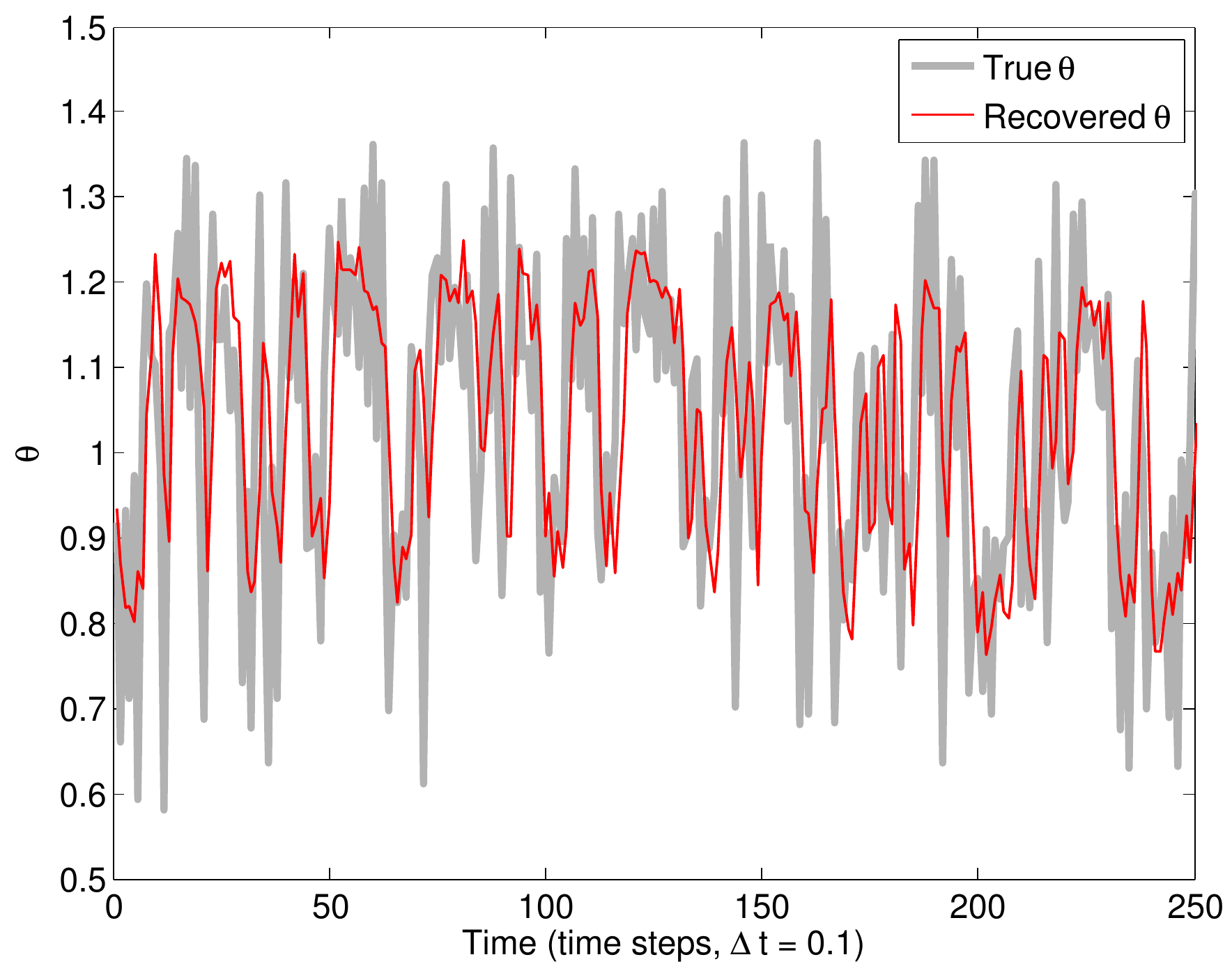}
\includegraphics[width=0.45\textwidth]{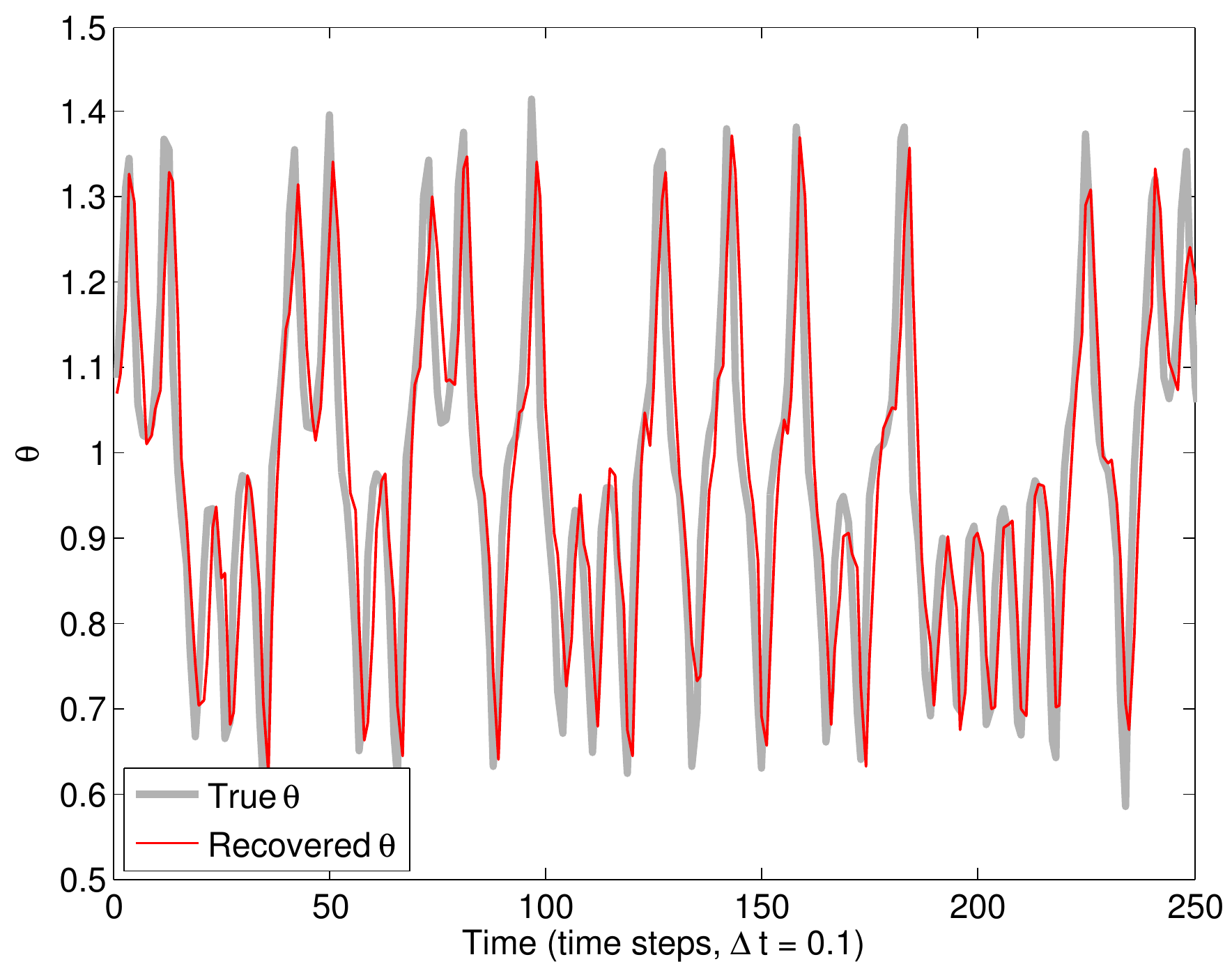}
\includegraphics[width=0.45\textwidth]{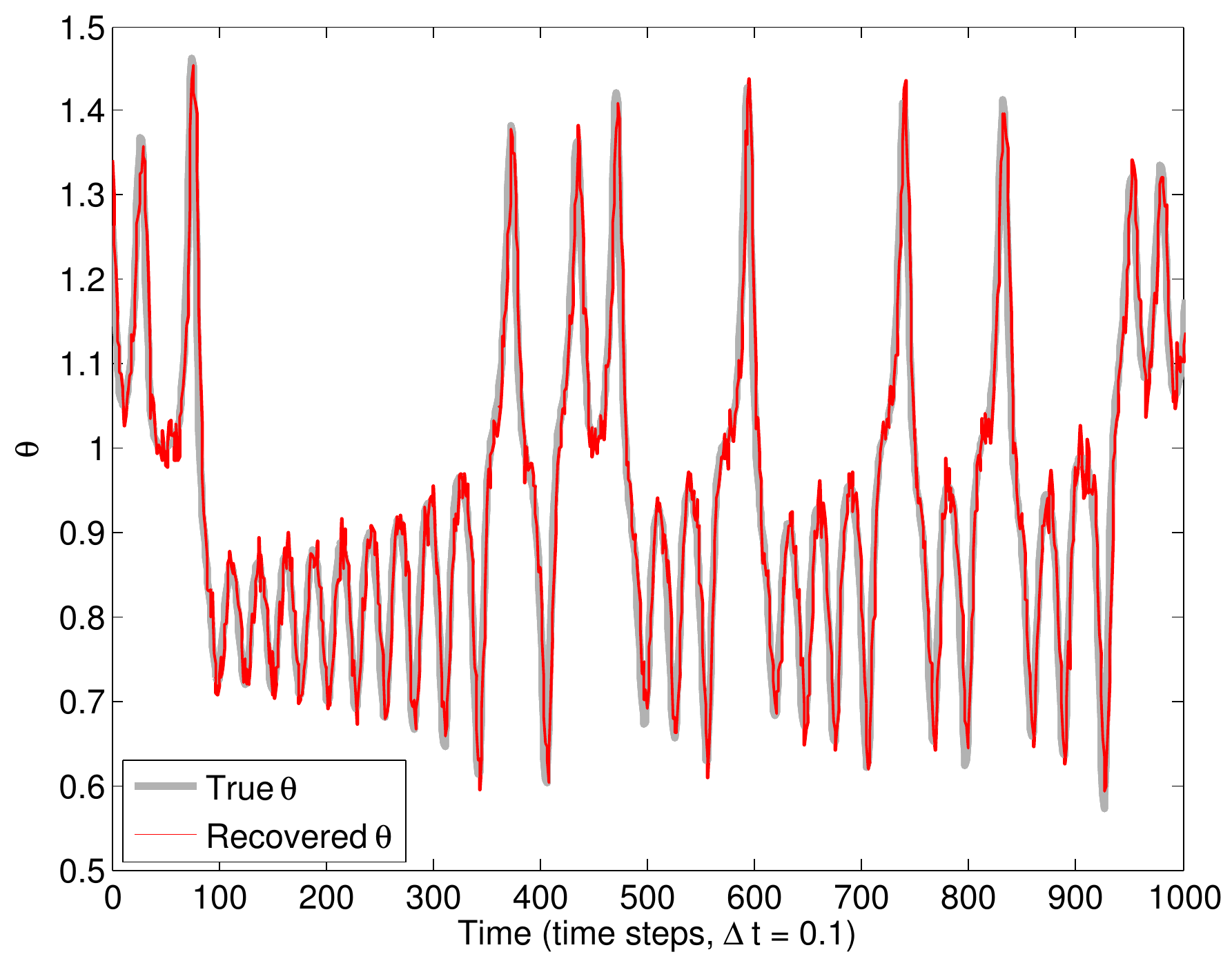}
\caption{\label{recoveringtheta} Comparison the true time series of $\theta$ and the recovered time series for $\epsilon =0.25$ (left), $\epsilon=1$ (middle), and $\epsilon =4$ (right).}
\end{figure}

Once the value of $Q_{\theta\theta}$ is found from the adaptive EnKF, we use a standard EnKF to solve the following filtering problem,
\begin{align}\label{estimationfilter} 
\dot x &= f(x,\theta)  \nonumber \\
 \dot \theta &= \sqrt{Q_{\theta\theta}}\dot W \\
y^o &= y + \eta = h(x) + \eta, \nonumber
\end{align}
where the goal is to recover the optimal estimates of $(x,\theta)$ at each time $t_i$, given all the observations $y^o(t_j)$ for $j\leq i$.  Note that in the examples in this paper there is no stochastic forcing on the $x$ variables, however this approach would allow an additive white noise forcing term of the form $\dot x = f(x,\theta) + \sqrt{Q_{xx}} \dot W_x$.  Furthermore, we will only consider examples where the observation function $h$ has no explicit dependence on $\theta$ since having direct observations of $\theta$ would typically make recovering $\theta$ less challenging.  In Figure \ref{recoveringtheta} we show the results of applying this algorithm for the model \eqref{l96l63} using an unscented ensemble Kalman filter \citep{julier2004unscented} with 82 ensemble members (the filter equations used here are the same as those in Section \ref{filter} for the parametric model).  

In this example, we use a time series of 5000 noisy observations $y^o(t_i)$ with $\Delta t = t_{i+1}-t_i = 0.1$, the observation function is $h(x,\theta)=x$, and the observation noise variance is $R=\frac{1}{8} I_{40}$. In Figure \ref{recoveringtheta} we show the true time series of $\theta$ and the recovered time series for $\epsilon=0.25, 1, 4$. In each case, the the true variance of $\theta$ is $0.039$, and the estimated variances are $0.0192$, $0.0193$, and $0.0066$ for $\epsilon = 0.25,1,4$ respectively. When the Lorenz-63 system evolves more slowly, as in the cases $\epsilon=1,4$, a very accurate recovery is possible. In contrast, when $\epsilon=0.25$, the Lorenz-63 system evolves quickly relative to the observations $y$ of the state $x$, which makes clean recovery challenging, however the recovered time series still has a high correlation with the true dynamics. Obviously, one would get worse recovery when observation function $h$ is a more complicated operator (for example, if the observations of the state $x$ are sparse) or when the observation noise covariance $R$ is larger. While these issues can be important in practice, we omit exploring them here since it is more related to the accuracy of the primary filter (EnKF) and our assumption is that $\theta$ can be recovered with an appropriate ensemble filtering method from the existing ensemble forecasting infrastructure.   

In Section \ref{filter} we will compare the filtering and forecasting skill of the semiparameteric approach to the various parametric models introduced in Section \ref{problem} for these three time scales ($\epsilon=0.25,1,4$). As we will see, even when the parameters evolve quickly the recovered data set is still useful in overcoming the model error.

\section{Semiparametric filtering}\label{filter}

\begin{figure*}
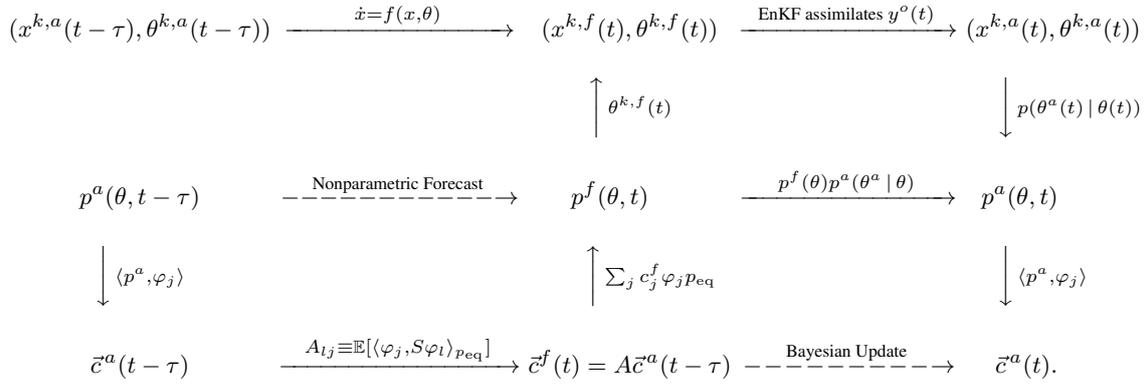

\[ \arraycolsep=1.4pt\begin{array}{ccccc}
(x^{k,a}(t-\tau),\theta^{k,a}(t-\tau))\hspace{0pt}&\xrightarrow{\ \ \ \ \ \ \ \ \ \ \dot x = f(x,\theta) \ \ \ \ \ \ \ \ \ \ }&(x^{k,f}(t),\theta^{k,f}(t))&\xrightarrow{\ \ \textup{EnKF assimilates } y^o(t) \ \ }&(x^{k,a}(t),\theta^{k,a}(t)) \\ \\
&&\left\uparrow\rule{0cm}{.5cm}\right. \scriptstyle{\theta^{k,f}(t)}&&\hspace{12pt}\left\downarrow\rule{0cm}{.5cm}\right. \scriptstyle{p(\theta^a(t) \, | \, \theta(t))} \\ \\
p^a(\theta,t-\tau) \hspace{0pt}&\xdashrightarrow{\ \ \ \ \textup{Nonparametric Forecast} \ \ \ \ }& \hspace{-15pt}p^f(\theta,t) &\xrightarrow{\ \ \ \ \ p^f(\theta)p^a(\theta^a \, | \, \theta) \ \ \ \ \ }& \hspace{-25pt}p^a(\theta,t) \\ \\
\left\downarrow\rule{0cm}{.5cm}\right.   \scriptstyle{\left<p^a,\varphi_j\right>}&& \hspace{15pt}\left\uparrow\rule{0cm}{.5cm}\right. \scriptstyle{\sum_j c_j^f \varphi_j \peq}&&\hspace{-8pt}\left\downarrow\rule{0cm}{.5cm}\right.   \scriptstyle{\left<p^a,\varphi_j\right>}\\ \\
\vec c^{\,a}(t-\tau) \hspace{0pt}&\xrightarrow{\ \ \ A_{lj} \equiv \mathbb{E}[\langle \varphi_j, S\varphi_l\rangle_{\peq}] \ \ \ }&  \vec c^f(t) = A  \vec c^{\,a}(t-\tau) &\xdashrightarrow{\ \ \ \ \ \ \textup{Bayesian Update} \ \ \ \ \ \ }&  \hspace{-20pt}\vec c^{\,a}(t).
\end{array} \]
\caption{\label{filter diagram} Diagram summarizing the semiparmetric filtering algorithm.}
\end{figure*}

In this section we assume that a training data set has already been recovered using the method of Section \ref{findp} and our goal is to estimate the current state, $x(t)$, and the posterior density $p(\theta,t)$ at some time $t$ given a sequence of noisy observations $y^o(s)$ for $s\leq t$, which will serve as the initial conditions for the forecasting problem of Section \ref{forecast}. Recall that the nonparametric forecasting procedure involves writing a density in the basis $\{\varphi_j\}$ and propagating the coefficients $c_j$ forward in time with a linear map $\hat{A}$, where $\hat{A}$ and $\varphi_j$ are built using the time series extracted by the method in Section~\ref{findp}.  Therefore, we are actually interested in the coefficients $c_j(t)$ which describe the conditional density $p(\theta(t) \, | \, y^o(s), s\leq t)$ written in the basis $\varphi_j$.  In other words, given all of the observations up to and including the current observation, we would like to find the optimal description of the current state. To solve this problem, we need to combine our nonparametric forecast model with the parametric filter to form a semiparametric filter.  

The goal of semiparametric filtering is to combine the parametric model, $f$, for the evolution of $x$ with a nonparametric model for the evolution of $\theta$ to determine the filtered state $p(x(t),\theta(t) \, | \, y^o(s), s\leq t)$ from the previous filtered state $p(x(t-\tau),\theta(t-\tau) \, | \, y^o(s), s\leq t-\tau)$.  The algorithm introduced here will follow the standard \emph{forecast-assimilate} paradigm for filtering, in which the \emph{forecasting} step will be to perform a one step forecast to determine $p(x(t),\theta(t) \, |\, y^o(s), s\leq t-\tau)$ and the \emph{assimilation} step uses the observation $y^o(t)$ to find the final density $p(x(t),\theta(t) \, | \, y^o(s), s\leq t)$.  

We introduce the notation $x^{k,a}$ and $\theta^{k,a}$ to denote an ensemble of states $x$ and parameters $\theta$. The superscript $k$ denotes the $k$-th ensemble member, and the superscript `a' stands for analysis, indicating that this ensemble has assimilated all the information up to the current time, including the current observation. Similarly, $x^{k,f}$ and $\theta^{k,f}$ denote the forecast (also known as `prior') ensemble, as indicated by the superscript `f', which account for the observations up to the previous time step. To be consistent, we also use same superscripts `a' and `f' to denote the nonparametric analysis density $p^a(\theta,t)$ and the nonparametric forecast density $p^f(\theta,t)$ as well as the associated coefficients $\vec c^{\,a}$ and $\vec c^f$ respectively.

We overview the semiparametric filtering algorithm with the diagram shown in Figure \ref{filter diagram}. As shown in Figure \ref{filter diagram}, information will flow up from the density to the ensemble in the forecast step and back down from the ensemble to the density in the assimilation step.  The basic idea of the semiparametric filter is to apply the EnKF to estimate a Gaussian approximation to the posterior $p(x,\theta(t) \, | \, y^o(s), s\leq t)$ and then to treat the parametric filter estimate $\theta^a(t)$ as a noisy observation which we feed into a secondary nonparametric filter.  The forecast step for $\theta$ can then be performed using the nonparametric model, and the first two statistics of this forecast are substituted into the Kalman filter prior statistics for $\theta$.  

To make this procedure precise, assume that we are given an estimate of the state $(x^a(t-\tau),\theta^a(t-\tau))$ along with a covariance matrix $C^a(t-\tau)$ of the form \eqref{covmat} and coefficients $\vec c^{\,a}(t-\tau)$ which represent the current posterior density $p^a(\theta,t-\tau)$ given all the observations $y^o(s)$ for $s\leq t-\tau$.  The filtering procedure is iterative, so given this information, it will assimilate the next observation $y^o(t)$ and produce $(x^a(t),\theta^a(t))$, $C^a(t)$ and $\vec c^{\,a}(t)$.  

To perform the forecast step of the semiparametric filter, we first form an ensemble $\{(x^{k,a}(t-\tau),\theta^{k,a}(t-\tau))\}_{k=1}^K$ of states which match the statistics $(x^a(t-\tau),\theta^a(t-\tau))$ and $C^a(t-\tau)$. In all the examples in this paper we use the unscented square root ensemble \citep{julier2004unscented} as described in Section \ref{perfectexample}. Next, we use the model \eqref{model} to integrate this ensemble forward in time to find the forecast ensemble $\{x^{k,f}(t)\}$ (also known as the background ensemble).  In order to update the statistics of $\theta$, we run a single diffusion forecast step $\vec{c}^{\,f}(t) = \hat A \vec{c}^{\,a}(t-\tau)$.  Using the ensemble $\{x^{k,f}(t)\}$ and the coefficients $\vec{c}^{\,f}(t)$, we compute the forecast statistics (also known as the prior statistics or background statistics) by mixing the parametric and nonparametric information as follows,
\begin{align}\label{priorstats} 
x^f(t) &= \frac{1}{K}\sum_{k=1}^K x^{k,f}(t) \\
C^f_{xx}(t) &=  \frac{1}{K-1}\sum_{k=1}^K (x^k(t)-x^f(t))(x^k(t)-x^f(t))^\top \nonumber \\
\theta^a(t-\tau) &= \frac{1}{K} \sum_{k=1}^K \theta^{k,a}(t-\tau) \nonumber  \\
 C^f_{x\theta}(t) &=  C^f_{\theta x}(t)^\top \nonumber \\
 &= \sum_{k=1}^K \frac{(x^{k,f}(t)-x^f(t))(\theta^{k,a}(t-\tau)-\theta^a(t-\tau) )^\top}{K-1} \nonumber  \\
 p^f(\theta_i,t)  &=\sum_{j=1}^{M}   c_j(t)  \varphi_j(\theta_i) \peq(\theta_i) \nonumber \\
 \theta^f(t) &= \int_{\theta\in\mathcal{M}} \theta p^f(\theta,t) \approx \frac{1}{N}\sum_{i=1}^N \theta_i p^f(\theta_i,t)  \nonumber \\
 C^f_{\theta\theta}(t) &= \int_{\theta\in\mathcal{M}} (\theta-\theta^f(t))(\theta-\theta^f(t))^\top p^f(\theta,t) \nonumber \\ &\approx \frac{1}{N}\sum_{i=1}^N (\theta_i-\theta^f(t))(\theta_i-\theta^f(t))^\top p^f(\theta_i,t)  \nonumber \\
 C^f(t) &= \left(\begin{array}{cc}  C^f_{xx}(t) &  C^f_{x\theta}(t) \\  C^f_{\theta x}(t) &  C^f_{\theta \theta}(t) \nonumber \end{array}\right).
\end{align}
Notice that we use the ensemble to compute the statistics of $x$ and the correlation between $x$ and $\theta$, while we use the nonparametric model to compute the statistics of $\theta$. Moreover, the cross covariance terms $C_{x\theta}^f(t) = C_{\theta x}^f(t)^\top$ simply use the analysis ensemble $\theta^{k,a}(t-\tau)$ from the previous time step. This is required since we do not integrate the individual initial conditions $\theta^{k,a}$, but rather the full density.  Using the semiparametric forecast mean $z^f(t) \equiv (x^f(t),\theta^f(t))$ and covariance $C^f(t)$, we now resample the forecast ensemble $z^{k,f}(t) = (x^{k,f}(t),\theta^{k,f}(t))$, (we use a square root ensemble here) and for convenience we will use the same symbols for the resampled forecast ensemble. Naively, one might hope to simply apply the rejection sampling strategy to generate $\theta^{k,f}$, and keep using the ensemble $x^{k,f}$ from the ensemble forecast without resampling; however, this strategy does not preserve the cross-covariance structure between $x$ and $\theta$ from the previous filter step. We attempted this alternative strategy and found that it significantly degraded the filter performance compared to resampling the ensemble $z^{k,f}(t)$ using \eqref{priorstats}. With the newly resampled ensemble, we form the observation ensemble $y^{k,f}(t) = h(x^{k,f}(t),\theta^{k,f}(t))$ by applying the observation function to the reformed ensemble.  Notice that in general the observation function may depend on both the state $x$ and the parameters $\theta$, but in the examples below we will only consider the more difficult case where the observation function only depends on $x$ (so that $\theta$ is not directly observed).  

With the forecast step completed, we now perform the assimilation step, which will consist of an EnKF assimilation step and a secondary nonparametric assimilation step.  We first perform the EnKF update in order to assimilate the information given by observations, $y^o(t)$, at time $t$, with the standard equations,
\begin{align}\label{update}
y^f(t) &= \frac{1}{K}\sum_{k=1}^K y^{k,f}(t)  \\
z^f(t) &= \frac{1}{K}\sum_{k=1}^K z^{k,f}(t)  \nonumber \\
C_{zy}(t) &= \frac{1}{K}\sum_{k=1}^K (z^{k,f}(t)-z^f(t))(y^{k,f}(t)-y^f(t))^\top  \nonumber \\
C_{yy}(t) &=  \frac{1}{K}\sum_{k=1}^K (y^{k,f}(t)-y^f(t))(y^{k,f}(t)-y^f(t))^\top \nonumber \\
K(t) &= C_{zy}(t)C_{yy}(t)^{-1}  \nonumber \\
    z^{a}(t) &= (x^a(t),\theta^a(t)) =  z^f(t)+K(t)(y^o(t)-y^f(t))  \nonumber \\
    C^a(t) &= C^f(t) - K(t) C_{yy}(t) K(t)^\top.  \nonumber
\end{align}
The ensemble Kalman update \eqref{update} yields the analysis mean $(x^a(t),\theta^a(t))$ and covariance $C^a(t)$ from the parametric model, and it remains to update the nonparametric model coefficients $\vec{c}^{\,f}(t)$ in order to find the analysis coefficients $\vec{c}^{\,a}(t)$ which are required in order to iterate the filter.  Notice that if we do not update these coefficients to incorporate information from $y^o(t)$, they will evolve to represent the equilibrium density $\peq$, and the forecast $\theta^f$ will eventually be constant, so this secondary assimilation step is crucial.  

In order to update $\vec{c}^{\,f}(t)$, we will take the posterior estimate $\theta^a(t)$ from the EnKF, and treat it as a noisy observation, which is reasonable since it is only an estimate of the true state, and the filter also gives us an error estimate, namely the analysis covariance, $C^a_{\theta\theta}(t)$.  If the observations were continuous in time, we could filter this observation using the Zakai equation projected onto the basis $\{\varphi_j\}$, and this was the approach taken in \cite{BH14UQ}.  However, since the observation time may be long, we will apply a Bayesian update which was introduced in \cite{BH15PHYSD}. Explicitly, we want to find the posterior density $p^a(\theta,t) = p(\theta,t \, | \, y^o(s), s\leq t)$ by combining the forecast density $p^f(\theta,t) = p(\theta,t \, | \, y^o(s), s < t) = \sum_{j=1}^M c^f_j(t)\varphi_j(\theta)\peq(\theta)$ from the nonparametric model with the EnKF analysis estimate $\theta^a(t)$. Since the ensemble Kalman filter assumes the analysis solutions to be Gaussian, we have the following marginal analysis density for $\theta$, 
\begin{align}
p(\theta | y^o(t)) \propto \exp(-\frac{1}{2}||\theta-\theta^a(t)||_{C^a_{\theta\theta}}^2).\label{likelihood}
\end{align}
We will treat this Gaussian analysis density as a likelihood function, that is, we define the likelihood function $p(\theta^a(t)\, |\,\theta) = p(\theta \, | \, y^o(t))$. We now use this likelihood function in the following Bayesian assimilation scheme,
\begin{align}\label{bayes} p^a(\theta,t) &\propto  p^f(\theta,t)p(\theta^a(t)\,|\,\theta) \nonumber \\ &\propto \sum_{j=1}^M c^f_j(t)\varphi_j(\theta)\peq(\theta)\exp(-\frac{1}{2}||\theta-\theta^a(t)||_{C^a_{\theta\theta}}^2). \end{align}
By evaluating \eqref{bayes} at each data point $\theta_i$ from the training data set, we can find $p^a(\theta,t)$ up to a normalization factor which we compute with \eqref{normalizationfactor}.  We can then compute the updated coefficients $c_l^a(t)$ by projecting the analysis density $p^a(\theta,t)$ onto the basis $\varphi_l$ by computing,
\begin{align}\label{nonparamupdate} c_l^a(t) = \left<p^a(\theta,t),\varphi_l \right> \approx \frac{1}{N}\sum_{i=1}^N \frac{p^a(\theta,t)}{\peq(\theta_i)}\varphi_l(\theta_i).
\end{align}
With the updated coefficients $\vec c^{\,a}(t)$ from \eqref{nonparamupdate} we are now ready to perform the next filter step.  Moreover, at this point we can apply the semiparametric forecasting algorithm of Section \ref{forecast} to the current state, represented by $x^a(t)$, $C^a_{xx}(t)$ and $\vec{c}^{\,a}(t)$ to estimate the future densities $p(\theta, t+m\tau \, | \, y^o(s), s\leq t)$ at future times $t+m\tau$, given only the information up to the current time, $t$.  

Initializing the mean $(x^a(0),\theta^a(0))$ and covariance $C^a(0)$, where time zero simply represents the first observation one desires to assimilate, is a standard problem in filtering and in this paper we will use the true initial state $(x(0),\theta(0))$ along with an empirically chosen diagonal covariance matrix with $C_{xx}^a = I_{40}$ and $C_{\theta\theta}^a=Q_{\theta\theta}$.  To initialize the coefficients $\vec{c}^{\,a}(0)$ we will use a non-informative prior, meaning that the initial density $p^a(\theta,0)=\peq(\theta)$ so that $c_j^a(0) = \left<\peq,\varphi_j\right> \approx \frac{1}{N}\sum_{i=1}^N \varphi_j(\theta_i)$.

\subsection{Semiparametric filtering and forecasting a system with chaotic model error}\label{realexample}

In this example we apply the semiparametric filtering technique to the system \eqref{l96l63} for $\epsilon \in \{0.25,1,4\}$ given only the model $f(x,\theta)$ and a time series of 5000 noisy observations $y^o(t_i)$ with $\Delta t = t_{i+1}-t_i = 0.1$. Implicitly, this means that we are applying the complete procedure, including estimating the variance of the parameters $\theta$ as in Appendix \ref{QR}, recovering a time series of the parameters $\theta$ as in Section \ref{findp}, building the nonparametric model for $\theta$, including applying diffusion maps algorithm to obtain the basis $\varphi_j(\theta_i)$ and the forecasting matrix $\hat A$ as discussed in Section~\ref{nonparametricforecast}, implementing semiparametric filtering as described in Section~\ref{filter}, and semiparametric forecasting as described in Section \ref{forecast}. Note that we apply the time delay embedding to the recovered time series $\theta_i$ and use this embedded data set to build the nonparametric model, including the equilibrium density $\peq(\theta_i)$, the basis $\varphi_j(\theta_i)$ and the forecasting matrix $\hat A$. Finally, we apply the semiparametric filter to an out-of-sample collection of 1000 noisy observations $y^o(t_i)$ where $i=5101,...,6100$. At each step, using the filter analysis as the initial conditions for the semiparametric forecast, we compute the forecast for 50 steps of length $\tau=\Delta t=0.1$ (which is 25 model days using the convention of 1 model day = 0.2 model time units) and we compare the mean of the forecast to the true state $x$ at the corresponding lead times.  

\begin{figure}[htbp]
\centering
\includegraphics[width=0.45\textwidth]{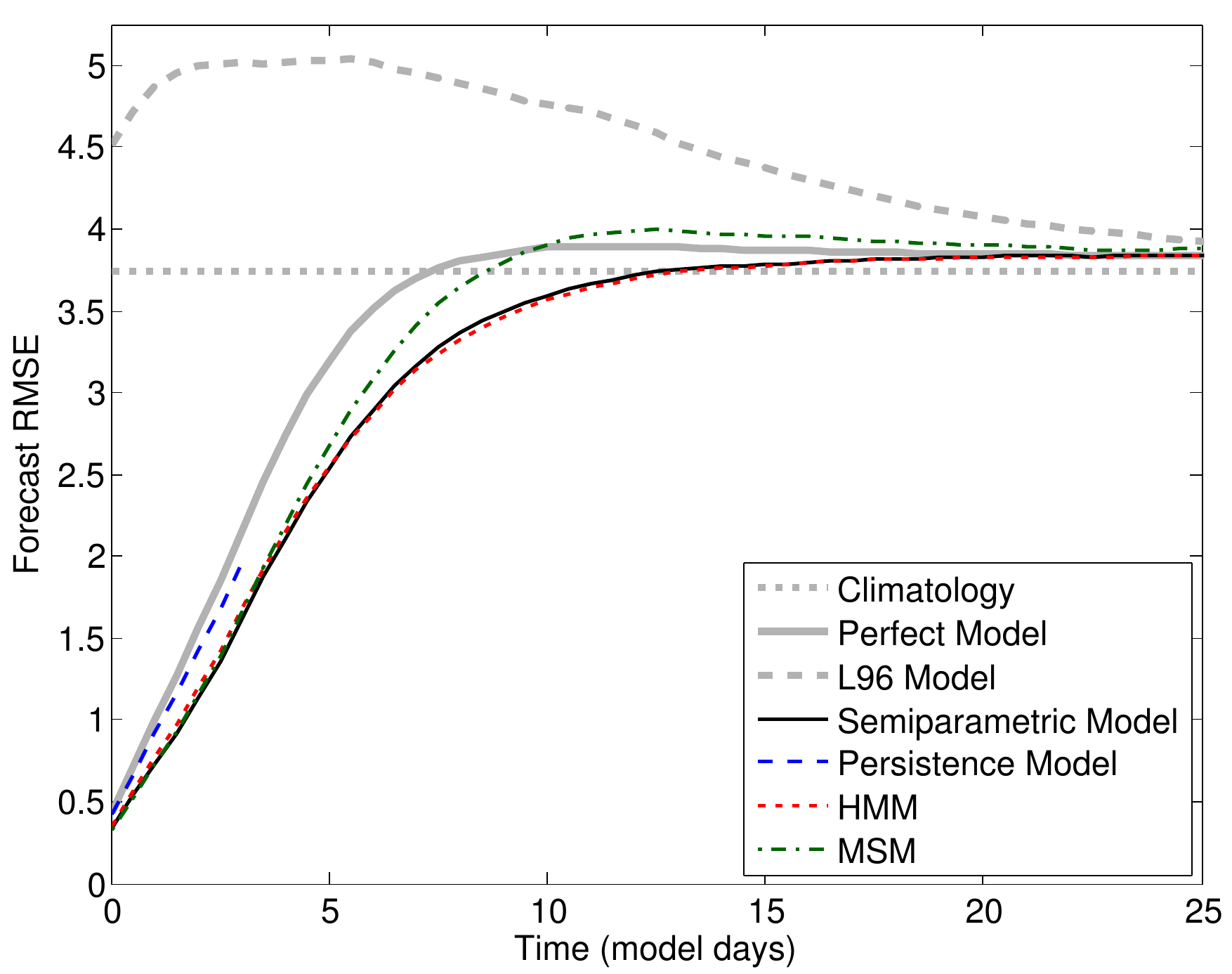}
\includegraphics[width=0.45\textwidth]{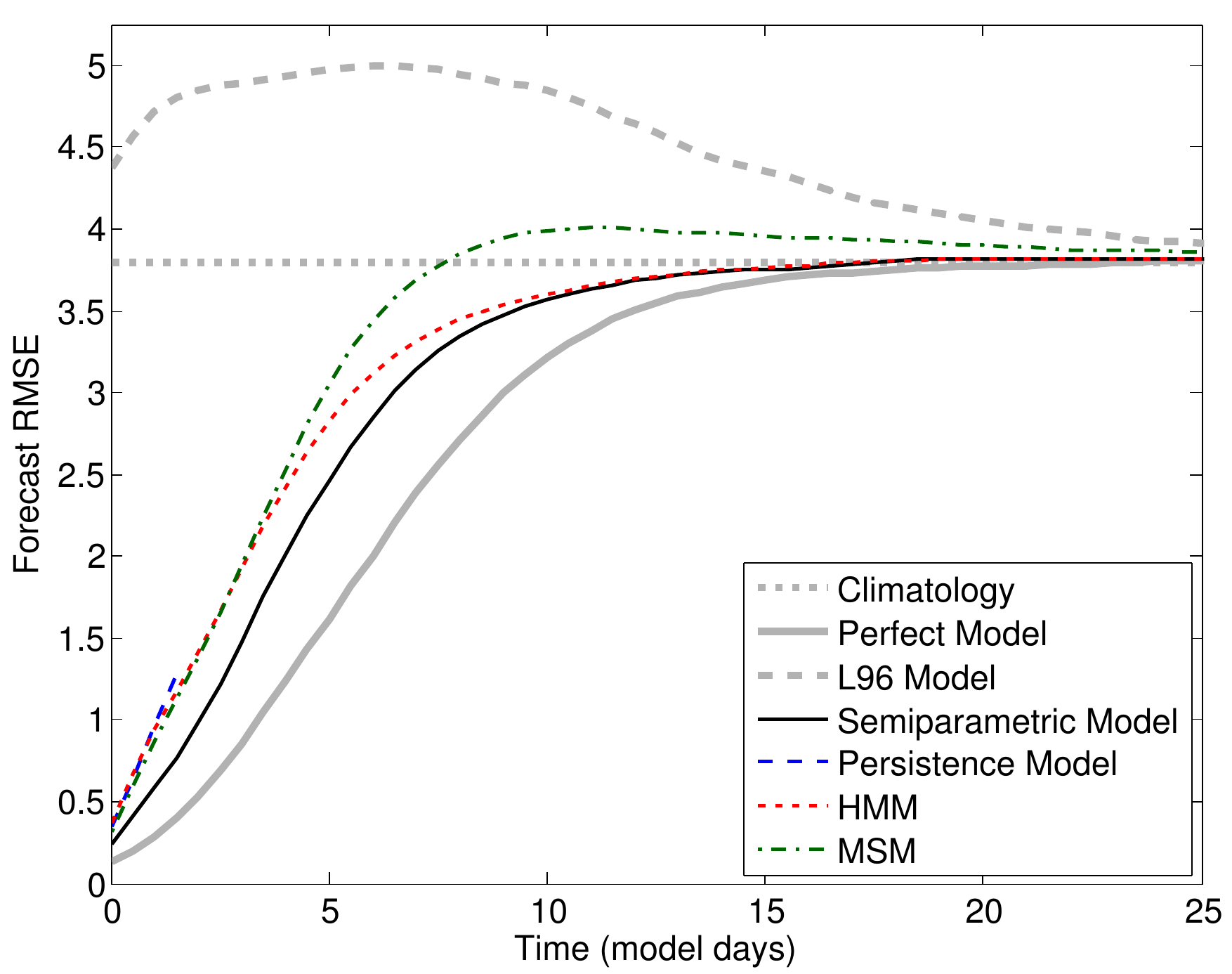}
\includegraphics[width=0.45\textwidth]{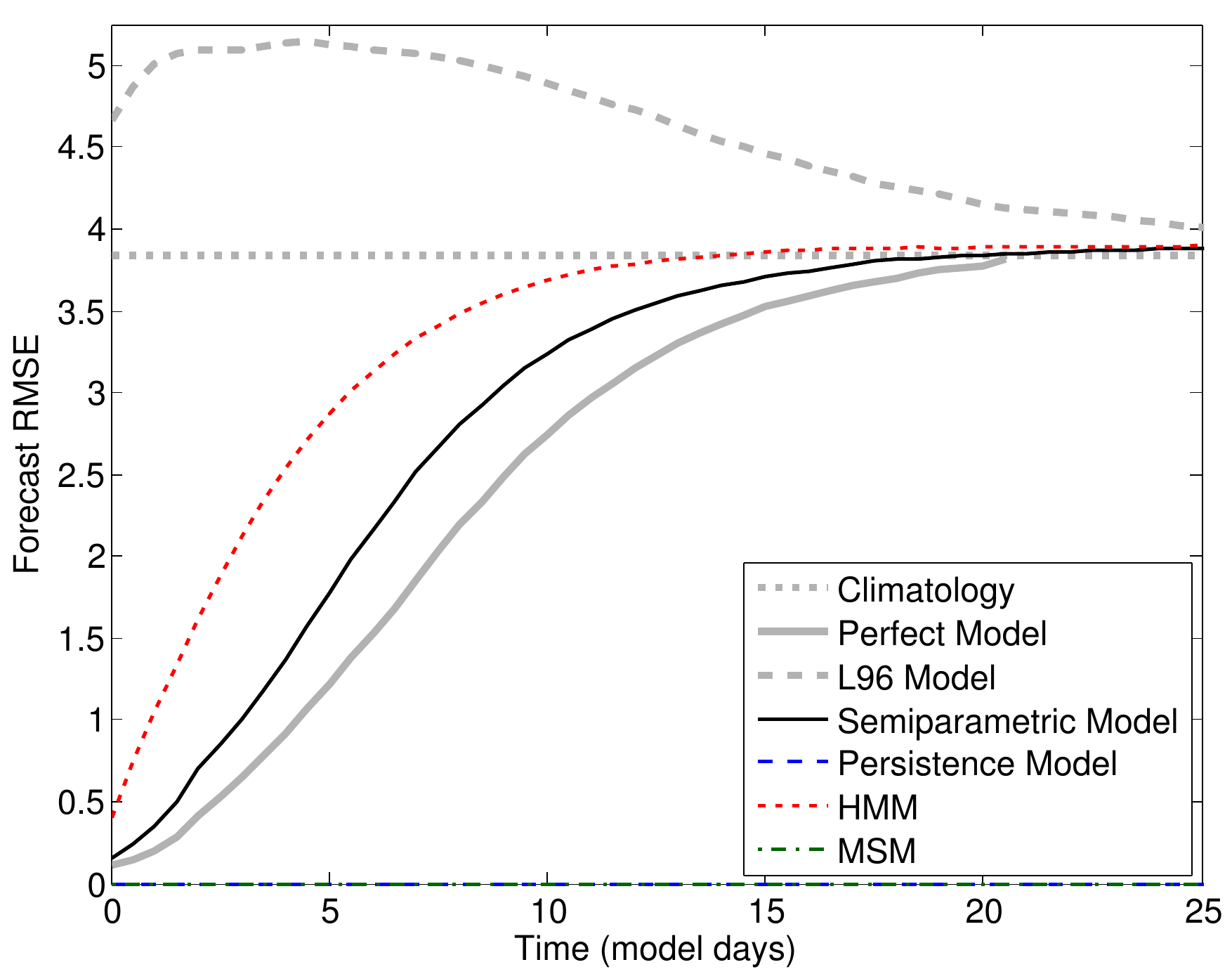}
\caption{\label{L96L63FilterEx} Comparison of forecasting methods for \eqref{l96l63} with recovered training data and filtered initial condition for $\epsilon = 0.25$ (left), $\epsilon=1$ (middle), and $\epsilon = 4$ (right).}
\end{figure}

 \begin{figure*}
\centering
\includegraphics[width=0.98\textwidth]{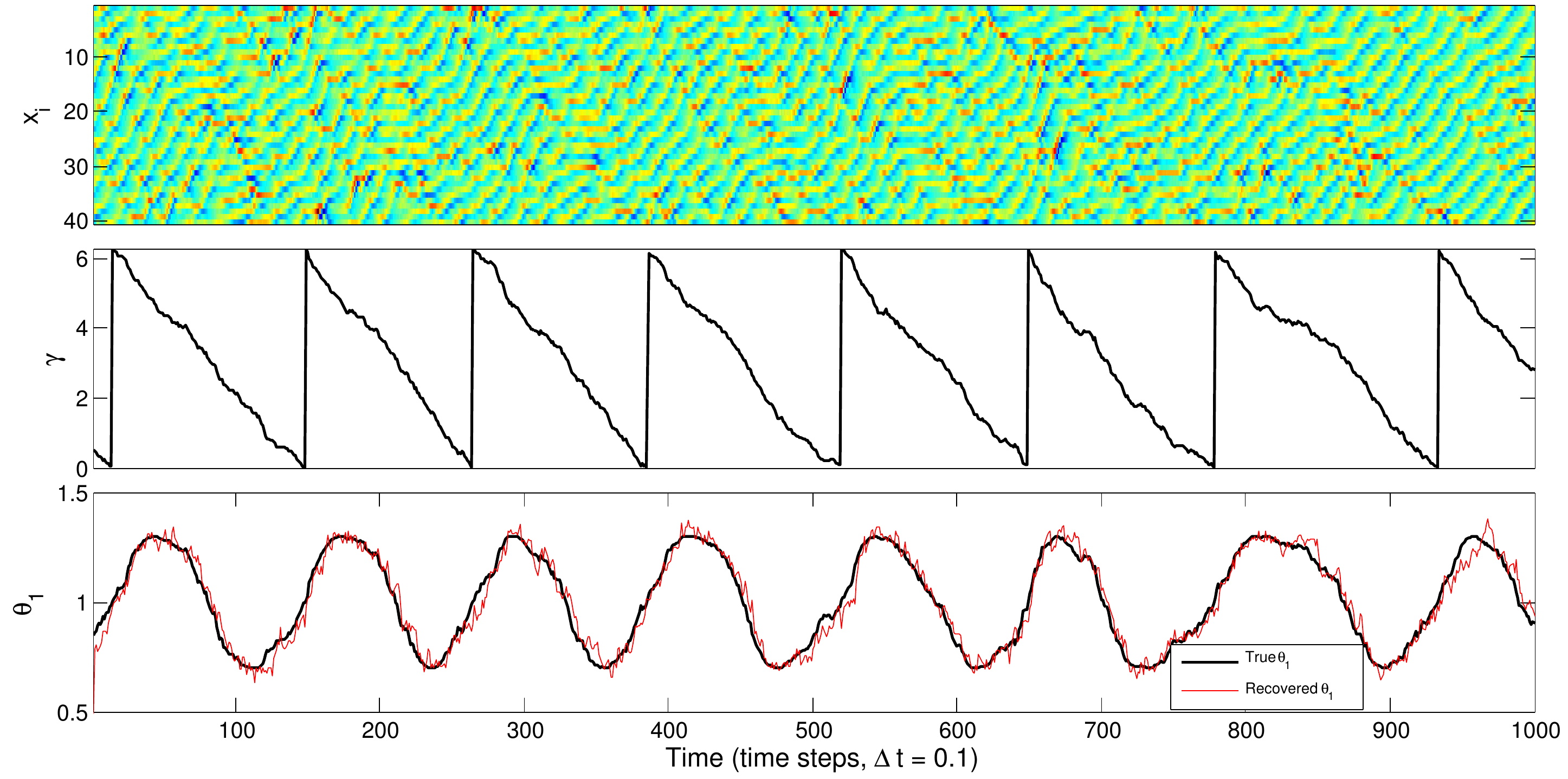}
\caption{\label{L96StochasticEx} Demonstrating the evolution of the system \eqref{l96stochastic} with $\epsilon=4$ (top) with the corresponding time series of the latent variable $\gamma$ (middle) and the first parameter $\theta_1$ (bottom, black curve) along with the recovered time series of $\theta_1$ estimated with the filter (bottom, red curve). }
\end{figure*}

In Figure \ref{L96L63FilterEx} we compare the semiparametric filter to alternative methods of correcting for the model error, all of which use the same training data set for $\theta_i$.  Notice that when the standard Lorenz-96 \eqref{l96} model is used without any modification, the ensemble Kalman filter cannot even estimate the initial state $x$, and the forecast is worse than the climatology of the true system. The perfect model produces the most accurate forecasts for larger $\epsilon=1, 4$. Notice that for $\epsilon=1$, the perfect model in Figure \ref{L96L63FilterEx} does not exhibit the bias as seen in the long term forecast of Figure \ref{L96L63Ex}. This is due to the ensemble from the filtered state having non-diagonal covariance structure which places the ensemble closer to the attractor of the true system \eqref{l96l63} and reduces the bias in the forecast. As shown in Figure \ref{L96L63FilterEx}, when $\epsilon=0.25$ the system \eqref{l96l63} is sufficiently stiff that the perfect model does not produce good filtered estimates and consequently the perfect model is outperformed by the alternative approaches.

The semiparametric filter and forecast give the next best forecasting skill and this approach is robust across the different values of $\epsilon$. When the Lorenz-63 system evolves more slowly ($\epsilon=1,4$), the semiparametric model has a significant advantage over all except for the perfect model. The next best approach is HMM, which samples randomly from the training data set \citep[see][for detailed implementation of the HMM in the data assimilation procedure]{kh:12}. This approach matches the semiparametric method when the Lorenz-63 system is very fast ($\epsilon=0.25$), which agrees with the theory of HMM in  \cite{HMM}, because when $\epsilon$ is small the Lorenz-63 system goes to the equilibrium density very quickly relative to the Lorenz-96 system which means that forecast model $\hat A$ cannot provide any additional information to the forecast. Finally, MSM performs quite well for small $\epsilon$, but the predictive skill degrades severely as $\epsilon$ increases (catastrophic divergence in the case of $\epsilon=4$). The persistence model forecast exhibits the catastrophic divergence for all values of $\epsilon$ since the model does not conserve energy.

\subsection{Semiparametric filtering and forecasting a system with stochastic model error}

In this example we apply the semiparametric filtering technique to the following system with four parameters $\theta =(\theta_1,...,\theta_4)$, each parameter $\theta_j$ is paired with the 10 state variables $x_i$ with $\textup{ceil}(i/10)=j$,
\begin{align}\label{l96stochastic}
f(x_i,\theta) &= \theta_{\textup{ceil}(i/10)} x_{i-1}x_{i+1}-x_{i-1}x_{i-2} - x_i + 6 \nonumber \\ 
\theta_j &= 1+(3/10)\sin(\gamma+(\pi/4)j) \\
\dot \gamma &=-(2-\sin(2\gamma)/2)/\epsilon + \sqrt{.1/\epsilon} \, \dot{W}, \nonumber 
\end{align}
 for $\epsilon \in \{0.25,1,4\}$.  Although the parameter space $\theta$ is four dimensional, these parameters actually lie on a one-dimensional domain $\mathcal{M}$ embedded in the ambient 4-dimensional space.  In this example, we set this one dimensional domain to be a circle with latent coordinate $\gamma$ which has a state dependent drift and a small amount of diffusion on the circle. We chose this example to demonstrate the semiparametric model on a higher-dimensional parameter space $\theta \in \mathbb{R}^4$ with stochastic evolution.  Since the parameters $\theta$ are actually constrained to a one-dimensional physical domain,  we expect to learn the model for the parameters with a very small training data set of noisy observations.  Finally, this example will demonstrate the strength semiparametric approach since it is nontrivial to represent the hidden latent variable $\gamma$ with a parametric ensemble forecast.  Given only the model $f(x,\theta)$ and a time series of 5000 noisy observations $y^o$ with $\Delta t = 0.1$, we apply the method of Appendix \ref{QR} to estimate the $4\times 4$ covariance matrix of the parameters $\theta$, and then we apply the method of Section \ref{findp} to recover the time series of $\theta$.  As an example, we show the recovered estimate of the time series of $\theta_1$ is compared to the true time series of $\theta_1$ in Figure \ref{L96StochasticEx} for $\epsilon=4$.

Even though the state $\gamma$ is fully observable from the four dimensional time series $\theta$, we still apply a time delay embedding to the recovered training time series with a single lag to smooth out the remaining noisy recovered time series. Notice in Figure \ref{L96StochasticEx} that the reconstruction of the true time series of parameters is somewhat noisy.  It was shown in \cite{BH15PHYSD}, applying the theory of \cite{DMDC}, that the delay embedding can be used to reduce the influence of the noise on the nonparametric model.  We note that while there is not a strong dependence on the number of lags used in the time delay embedding for this example, the theory introduced in \cite{DMDC} shows that as the number of delays increases, the geometry of the attractor becomes biased towards the stable components of the evolution, which are orthogonal to the noise.  

\begin{figure}[htbp]
\centering
\includegraphics[width=0.45\textwidth]{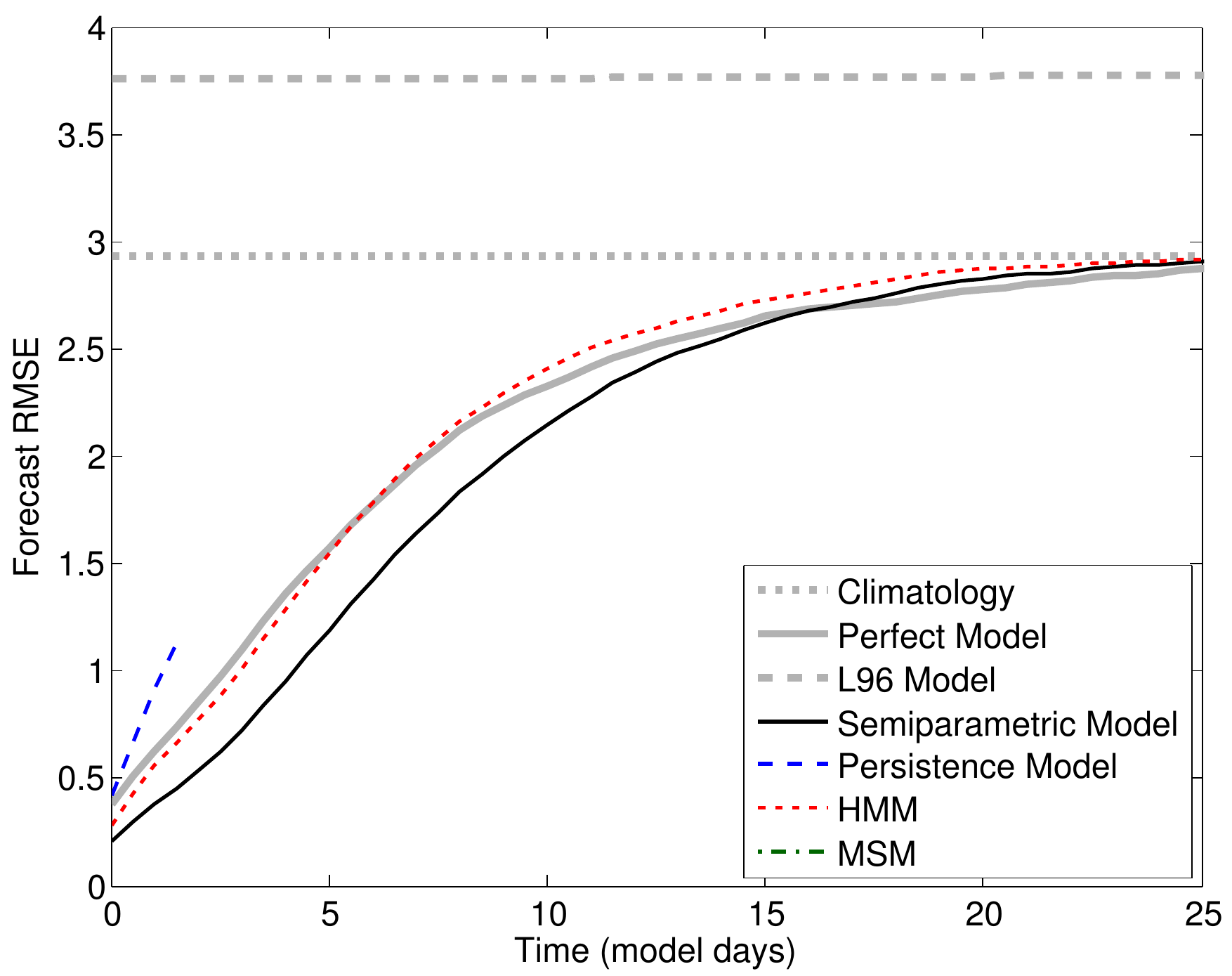}
\includegraphics[width=0.45\textwidth]{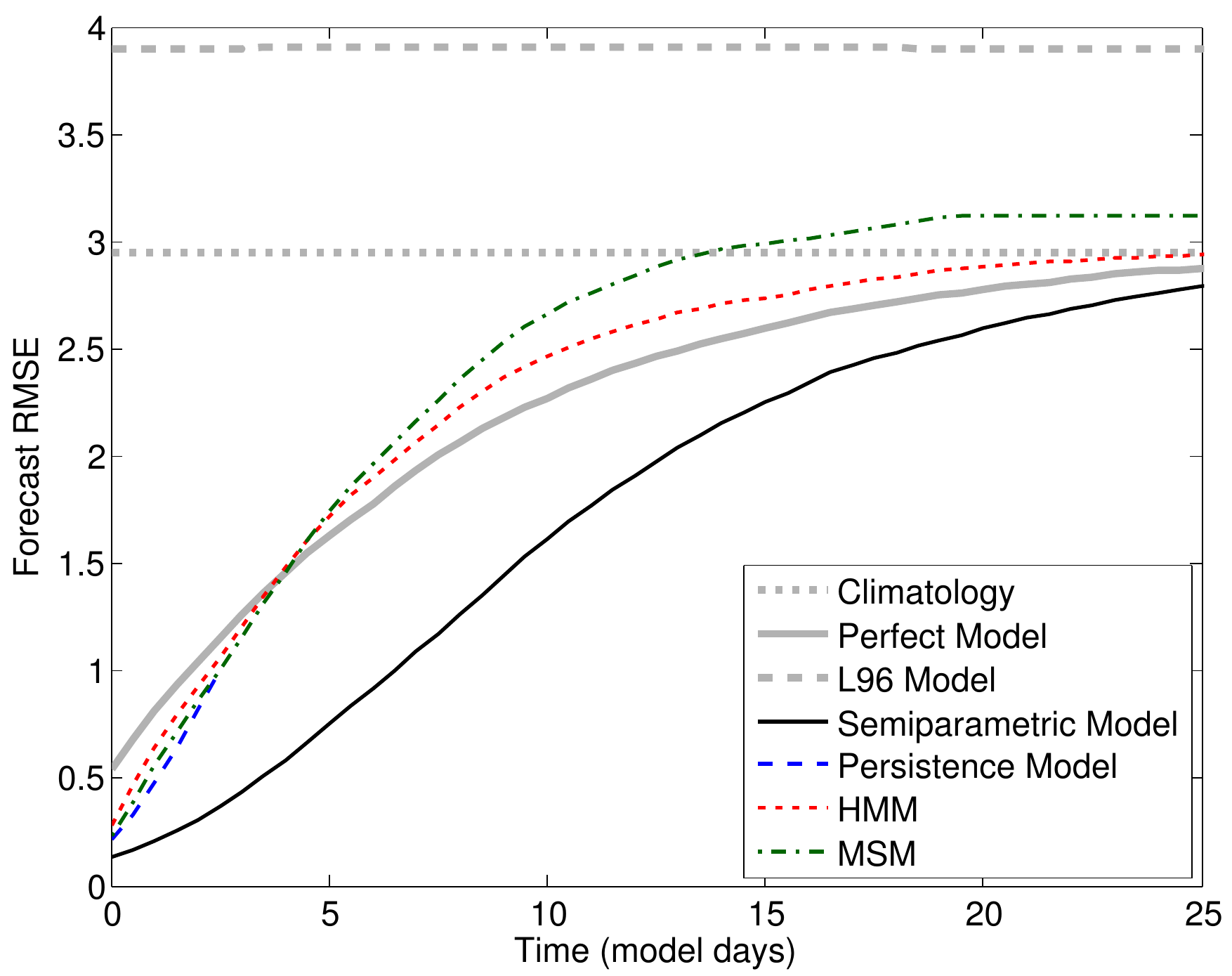}
\includegraphics[width=0.45\textwidth]{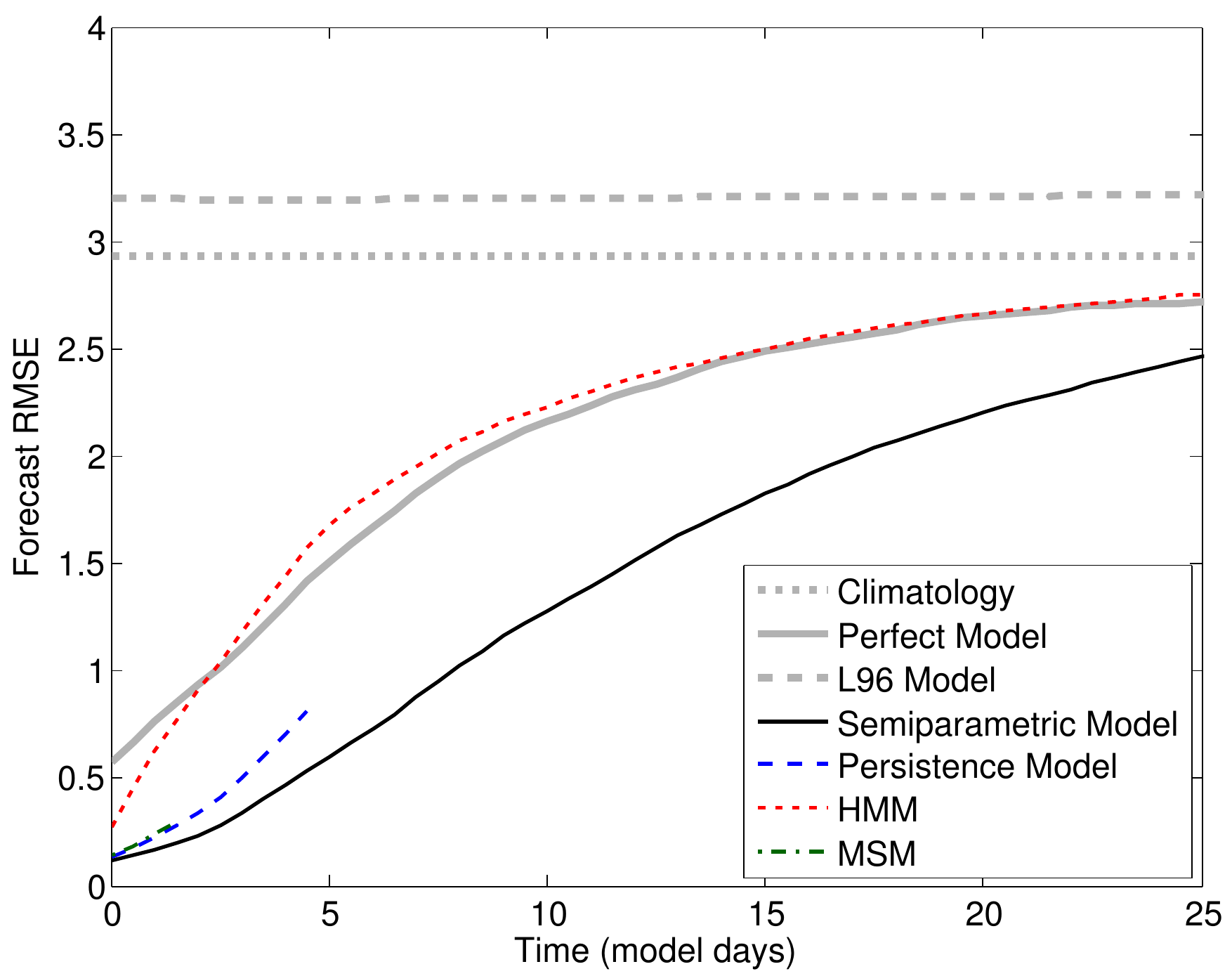}
\caption{\label{L96StochasticCompare} Comparison of forecasting methods for the L96-stochastic model with recovered training data and filtered initial condition for $\epsilon = 0.25$ (left), $\epsilon=1$ (middle), and $\epsilon = 4$ (right).}
\end{figure}

We learn the nonparametric model from 5000 recovered training data points of $\theta$, and apply the semiparametric filter to an out-of-sample collection of 1000 noisy observations $y^o(t_i)$ where $i=5101,...,6100$, computing the forecast up to the 50 step (5 model time unit, 25 model day) lead time as in the previous example.  In Figure \ref{L96StochasticCompare} we see that the nonparametric filter significantly outperforms the perfect model filter across all three time scales.  The poor performance of the perfect model is due to the small ensemble size, with only $82$ ensemble members unable to capture the evolution of the stochastic system \eqref{l96stochastic}.  Notice that any small perturbation of the latent variable $\gamma$ leads to significant variability in the high-dimensional chaotic evolution of $x$, so sufficiently sampling the uncertainty of $\theta$ is difficult with a small ensemble of only 82 paths.  In contrast, as previously noted, the semiparametric model does not produce paths of the evolution of $\gamma$, but rather samples independently from the forecast density at each forecast step.  Only when the system is very slow, $\epsilon=4$, is there a comparable filter, and in this case simply using the persistence model gives reasonable estimates.  However, while the persistence model gives a reasonably good forecast in the very short term for the case $\epsilon=4$, the long term forecast quickly degrades and in this case the forecast is unstable due to the dynamics \eqref{l96stochastic} being unstable when $\theta$ is fixed at a value far from $(1,1,1,1)$.  Attempting to use the standard Lorenz-96 system \eqref{l96} fails to filter the observations due to the model error, and while both the perfect model and the HMM give stable filters and forecasts they are significantly outperformed by the semiparametric model.

\section{Conclusion}\label{summary}

The goal of semiparametric modeling is to blend the strengths of parametric and nonparametric models. In particular, parametric models are able to use prior knowledge to describe intrinsically high-dimensional phenomena, however they are vulnerable to model error arising from unresolved phenomena or truncated scales.  On the other hand, nonparametric models learn directly from the data and place very few prior assumptions on the data, however the data requirements grow exponentially in the intrinsic dimensionality of the problem.  

For example, if one assumes that the most of the low frequency variability of the atmospheric dynamics, such as the Rossby waves pattern or North Atlantic Oscillation, can be captured by parametric models (such as the quasi-geostrophic models) then intuitively, if the model error is low-dimensional, it can be captured by a nonparametric model. In this paper, we introduce a modeling framework which can blend these approaches. We test our approach on a simple example, in which model error is introduced into the Lorenz-96 model with a parameter governed by either the Lorenz-63 model or a one-dimensional stochastic model. These examples were chosen as challenging test cases for semiparametric modeling in the following sense: The advection term in \eqref{l96l63} or \eqref{l96stochastic} does not actually conserve energy due to the parameter $\theta$. Of course, the average value of $\theta$ in \eqref{l96l63} or \eqref{l96stochastic} is 1, so averaged over a large enough time window the L96-L63 advection does approximately conserve energy. As an analogy to a real physical problem, treating the L96-L63 model as the truth, one can imagine that a coarse physical model may have suggested using the Lorenz-96 model, which exactly conserves energy, while the true system does not.  The difference between these two systems is very subtle as shown by the spatio-temporal patterns in Figure \ref{L96L63comp}. However, this small model error has a dramatic effect on forecasting skill as shown in Figure \ref{L96L63Ex} and even more so on the filtering as shown in Figure \ref{L96L63FilterEx}. The stochastic example in \eqref{l96stochastic} is an even a more difficult test bed in the sense that model errors are introduced by four parameters whose intrinsic dynamics is only one-dimensional. One can imagine that a standard parametric modeling approach may end up over-fitting the one-dimensional dynamics with four separate parametric models. 

The examples discussed above suggest that coarse physical models can be significantly improved by the semiparametric modeling paradigm.  Rather than choosing arbitrary models for unknown phenomena, the modeler can build a parametric model which captures all the known phenomena, and fit the remaining unknown phenomena with nonparametric modeling.  Of course, if some knowledge of the parameters is available, one should first extend the parametric model as far as possible using real physical knowledge. Only when the physical knowledge is exhausted should the semiparametric modeling be applied, thereby reducing the dimensionality of the model error as much as possible before learning the nonparametric model.  Choosing an appropriate parametric form to encode all the existing physical knowledge is a significant challenge, and the semiparametric modeling approach will need to take advantage of the many advancements in this field to reduce the load on nonparametric model.

The algorithms introduced in this paper strive to be practical by seamlessly blending the nonparametric model into a standard existing framework for parametric filtering and forecasting.  However, there are two significant practical limitations of this framework. First, in order to extract a time series of the parameters (to be used for building the nonparametric model), one requires a minimum observability condition for the parameters, as in any standard inverse problem. Assuming that this condition is met, we foresee that improving the methodology for extracting the time series of the hidden parameters from noisy data is the key remaining challenge to lifting this framework to real applications. Indeed, the approach introduced in Section \ref{findp} and Appendix \ref{QR} is intended only as a proof-of-concept which shows that extracting an unobserved time series of parameters is possible.  Second, the evolution of the parameters $\dot \theta = g(\theta, \dot W)$ is assumed to be independent of the state variable $x$ of the parametric model.  This assumption is currently required because the nonparametric model of \cite{BGH14} assumes that the evolution represented by the times series is ergodic, and if $g$ were allowed to depend on $x$ the time series $\theta$ would not be ergodic by itself.  Of course, combining the state variables $(x,\theta)$ would typically yield an ergodic time series, however, this system would once again be high-dimensional and would not be accessible to a nonparametric model.  Similarly, if $g$ depended on $x$, the theory of Takens \cite{Takens,SYC} suggests that a time-delay embedding of $\theta$ would attempt to reconstruct the attractor of the full system $(x,\theta)$ which again would again be limited by the curse-of-dimensionality. The most probable avenue of overcoming this restriction is to note that the condition probability of $p(\theta,t | x(t))$ has an evolution independent of $x$.  We expect that extending the theory of \cite{BGH14} to conditional densities of this form will be the key to constructing semiparametric models where the evolution of the parameters is state dependent.

\acks
The research of J.H. is partially supported by the Office of Naval Research Grants N00014-11-1-0310, N00014-13-1-0797, MURI N00014-12-1-0912 and the National Science Foundation DMS-1317919. T. B. is supported under the ONR MURI grant N00014-12-1-0912.

\bibliographystyle{wileyqj}
\bibliography{VBbib}

\begin{thebibliography}{38}
\providecommand{\natexlab}[1]{#1}
\providecommand{\url}[1]{\texttt{#1}}
\providecommand{\urlprefix}{URL }
\expandafter\ifx\csname urlstyle\endcsname\relax
  \providecommand{\doi}[1]{doi:\discretionary{}{}{}#1}\else
  \providecommand{\doi}{doi:\discretionary{}{}{}\begingroup
  \urlstyle{rm}\Url}\fi

\bibitem[{Arnold \emph{et~al.}(2013)Arnold, Moroz and Palmer}]{amp:13}
Arnold HM, Moroz IM, Palmer TN. 2013. Stochastic parametrizations and model
  uncertainty in the lorenz 96 system. \emph{Philosophical Transactions of the
  Royal Society A: Mathematical, Physical and Engineering Sciences}
  \textbf{371}(1991).

\bibitem[{Berry \emph{et~al.}(2013)Berry, Cressman, Feren\v{c}ek and
  Sauer}]{DMDC}
Berry T, Cressman JR, Feren\v{c}ek ZG, Sauer T. 2013. Time-scale separation
  from diffusion-mapped delay coordinates. \emph{SIAM J. Appl. Dyn. Syst.}
  \textbf{12}: 618--649.

\bibitem[{Berry \emph{et~al.}(2015)Berry, Giannakis and Harlim}]{BGH14}
Berry T, Giannakis D, Harlim J. 2015. Nonparametric forecasting of
  low-dimensional dynamical systems. \emph{Physical Review Letters [in review]}
  \url{http://arxiv.org/abs/1411.5069}.

\bibitem[{Berry and Harlim(2014{\natexlab{a}})}]{BH:14}
Berry T, Harlim J. 2014{\natexlab{a}}. Linear theory for filtering nonlinear
  multi scale systems with model error. \emph{Proc. R. Soc. A}
  \textbf{470}(2167).

\bibitem[{Berry and Harlim(2014{\natexlab{b}})}]{BH14UQ}
Berry T, Harlim J. 2014{\natexlab{b}}. Nonparametric uncertainty quantification
  for stochastic gradient flows. \emph{SIAM/ASA J. Uncer. Quant. [in review]}
  \url{http://arxiv.org/abs/1407.6972}.

\bibitem[{Berry and Harlim(2015{\natexlab{a}})}]{BH15PHYSD}
Berry T, Harlim J. 2015{\natexlab{a}}. Forecasting turbulent modes with
  nonparametric diffusion models. \emph{submitted to Physical D}
  \url{http://arxiv.org/abs/1501.06848}.

\bibitem[{Berry and Harlim(2015{\natexlab{b}})}]{BH14VB}
Berry T, Harlim J. 2015{\natexlab{b}}. Variable bandwidth diffusion kernels.
  \emph{Applied and Computational Harmonic Analysis}
  \doi{http://dx.doi.org/10.1016/j.acha.2015.01.001}.

\bibitem[{Berry and Sauer(2013)}]{bs:13}
Berry T, Sauer T. 2013. {Adaptive ensemble Kalman filtering of nonlinear
  systems}. \emph{Tellus A} \textbf{65}: 20\,331.

\bibitem[{Coifman and Lafon(2006)}]{diffusion}
Coifman R, Lafon S. 2006. Diffusion maps. \emph{Appl. Comput. Harmon. Anal.}
  \textbf{21}: 5--30.

\bibitem[{E \emph{et~al.}(2007)E, Engquist, Li, Ren, E, Engquist, Li and
  Ren}]{HMM}
E W, Engquist B, Li X, Ren W, E W, Engquist B, Li X, Ren W. 2007. Heterogeneous
  multiscale methods: A review. \emph{Commun. Comput. Phys} \textbf{2}(3):
  367--450.

\bibitem[{Epstein(1969)}]{epstein:69}
Epstein E. 1969. {Stochastic dynamic prediction}. \emph{Tellus Ser. A}
  \textbf{21}: 739--759.

\bibitem[{Frenkel \emph{et~al.}(2012)Frenkel, Majda and Khouider}]{fmk:12}
Frenkel Y, Majda AJ, Khouider B. 2012. Using the stochastic multicloud model to
  improve tropical convective parameterization: A paradigm example.
  \emph{Journal of the Atmospheric Sciences} \doi{10.1175/JAS-D-11-0148.1}.

\bibitem[{Friedland(1969)}]{friedland:69}
Friedland B. 1969. {Treatment of bias in recursive filtering}. \emph{IEEE
  Trans. Automat. Contr.} \textbf{AC-14}: 359--367.

\bibitem[{Friedland(1982)}]{friedland:82}
Friedland B. 1982. {Estimating sudden changes of biases in linear dynamical
  systems}. \emph{IEEE Trans. Automat. Contr.} \textbf{AC-27}: 237--240.

\bibitem[{Giannakis and Majda(2011)}]{giannakisMajda}
Giannakis D, Majda AJ. 2011. Time series reconstruction via machine learning:
  Revealing decadal variability and intermittency in the north pacific sector
  of a coupled climate model. In: \emph{Conference on Intelligent Data
  Understanding (CIDU)}. Mountain View, California, pp. 107--117.

\bibitem[{Giannakis and Majda(2012)}]{GiannakisPNAS}
Giannakis D, Majda AJ. 2012. Nonlinear laplacian spectral analysis for time
  series with intermittency and low-frequency variability. \emph{Proc. Nat.
  Acad. Sci.} \textbf{109}(7): 2222--2227, \doi{10.1073/pnas.1118984109}.

\bibitem[{Grabowski(2001)}]{grabowski:01}
Grabowski W. 2001. {Coupling cloud processes with the large-scale dynamics
  using the Cloud-Resolving Convection Parameterization (CRCP)}. \emph{J.
  Atmos. Sci.} \textbf{58}: 978--996.

\bibitem[{H{\"a}rdle(2004)}]{semiparamBook}
H{\"a}rdle W. 2004. \emph{Nonparametric and semiparametric models}. Springer.

\bibitem[{Harlim \emph{et~al.}(2014)Harlim, Mahdi and Majda}]{hmm:14}
Harlim J, Mahdi A, Majda A. 2014. An ensemble kalman filter for statistical
  estimation of physics constrained nonlinear regression models. \emph{Journal
  of Computational Physics} \textbf{257, Part A}: 782 -- 812.

\bibitem[{Julier and Uhlmann(2004)}]{julier2004unscented}
Julier SJ, Uhlmann JK. 2004. Unscented filtering and nonlinear estimation.
  \emph{Proceedings of the IEEE} \textbf{92}(3): 401--422.

\bibitem[{Kalnay(2003)}]{kalnay:03}
Kalnay E. 2003. \emph{{Atmospheric modeling, data assimilation, and
  predictability}}. Cambridge University Press.

\bibitem[{Kang and Harlim(2012)}]{kh:12}
Kang E, Harlim J. 2012. Filtering partially observed multiscale systems with
  heterogeneous multiscale methods-based reduced climate models. \emph{Monthly
  Weather Review} \textbf{140}(3): 860--873.

\bibitem[{Katsoulakis \emph{et~al.}(2004{\natexlab{a}})Katsoulakis, Majda and
  Sopasakis}]{kms:04}
Katsoulakis M, Majda A, Sopasakis A. 2004{\natexlab{a}}. {Multiscale couplings
  in prototype hybrid deterministic/stochastic systems: Part I, Deterministic
  closures}. \emph{Comm. Math. Sci.} \textbf{2}(2): 255--294.

\bibitem[{Katsoulakis \emph{et~al.}(2004{\natexlab{b}})Katsoulakis, Majda and
  Sopasakis}]{kms:05}
Katsoulakis M, Majda A, Sopasakis A. 2004{\natexlab{b}}. {Multiscale couplings
  in prototype hybrid deterministic/stochastic systems: Part II, Stochastic
  closures}. \emph{Comm. Math. Sci.} \textbf{3}(3): 453--478.

\bibitem[{Leith(1974)}]{leith:74}
Leith CE. 1974. Theoretical skill of monte carlo forecasts. \emph{Monthly
  Weather Review} \textbf{102}(6): 409--418.

\bibitem[{Lorenz(1996)}]{lorenz:96}
Lorenz E. 1996. {Predictability - a problem partly solved}. In:
  \emph{Proceedings on predictability, held at ECMWF on 4-8 September 1995}.
  pp. 1--18.

\bibitem[{Lorenz(1963)}]{L63}
Lorenz EN. 1963. Deterministic nonperiodic flow. \emph{J. Atmos. Sci.}
  \textbf{20}(2): 130--141.

\bibitem[{Majda \emph{et~al.}(2010)Majda, Gershgorin and Yuan}]{mgy:10}
Majda A, Gershgorin B, Yuan Y. 2010. {Low frequency response and
  fluctuation-dissipation theorems: Theory and practice}. \emph{J. Atmos. Sci.}
  \textbf{67}: 1181--1201.

\bibitem[{Majda and Harlim(2013)}]{mh:13}
Majda A, Harlim J. 2013. {Physics constrained nonlinear regression models for
  time series.} \emph{Nonlinearity} \textbf{26}: 201--217.

\bibitem[{Majda \emph{et~al.}(1999)Majda, Timofeyev and
  Vanden-Eijnden}]{mtv:99}
Majda A, Timofeyev I, Vanden-Eijnden E. 1999. {Models for stochastic climate
  prediction}. \emph{Proc. Nat. Acad. Sci.} \textbf{96}: 15\,687--15\,691.

\bibitem[{Majda and Grooms(2013)}]{mg:13}
Majda AJ, Grooms I. 2013. New perspectives on superparameterization for
  geophysical turbulence. \emph{Journal of Computational Physics (in press)} .

\bibitem[{Mehra(1972)}]{mehra:72}
Mehra R. 1972. Approaches to adaptive filtering. \emph{Automatic Control, IEEE
  Transactions on} \textbf{17}(5): 693--698.

\bibitem[{Sauer \emph{et~al.}(1991{\natexlab{a}})Sauer, Yorke and
  Casdagli}]{embedology}
Sauer T, Yorke J, Casdagli M. 1991{\natexlab{a}}. Embedology. \emph{Journal of
  Statistical Physics} \textbf{65}(3-4): 579--616, \doi{10.1007/BF01053745},
  \urlprefix\url{http://dx.doi.org/10.1007/BF01053745}.

\bibitem[{Sauer \emph{et~al.}(1991{\natexlab{b}})Sauer, Yorke and
  Casdagli}]{SYC}
Sauer T, Yorke J, Casdagli M. 1991{\natexlab{b}}. Embedology. \emph{J. Stat.
  Phys.} \textbf{65}(3): 579--616.

\bibitem[{Stark(1999)}]{stark1}
Stark J. 1999. Delay embeddings for forced systems. {I}. {D}eterministic
  forcing. \emph{Journal of Nonlinear Science} \textbf{9}: 255--332,
  \urlprefix\url{http://dx.doi.org/10.1007/s003329900072}.

\bibitem[{Stark \emph{et~al.}(2003)Stark, Broomhead, Davies and Huke}]{stark2}
Stark J, Broomhead D, Davies M, Huke J. 2003. Delay embeddings for forced
  systems. {II}. {S}tochastic forcing. \emph{Journal of Nonlinear Science}
  \textbf{13}: 519--577,
  \urlprefix\url{http://dx.doi.org/10.1007/s00332-003-0534-4}.

\bibitem[{Takens(1981)}]{Takens}
Takens F. 1981. Detecting strange attractors in turbulence. In: \emph{{I}n:
  {D}ynamical Systems and Turbulence, {W}arwick, {E}ds. {R}and, {D}. and
  {Y}oung, {L}.-{S}.}, \emph{Lecture Notes in Mathematics}, vol. 898, Rand D,
  Young LS\ (eds), Springer Berlin / Heidelberg, ISBN 978-3-540-11171-9, pp.
  366--381, \urlprefix\url{http://dx.doi.org/10.1007/BFb0091924}.

\bibitem[{Zhen and Harlim(2014)}]{ZH14}
Zhen Y, Harlim J. 2014. Adaptive error covariance estimation methods for
  ensemble kalman filtering. \emph{Journal of Computational Physics [in
  review]} \url{http://arxiv.org/abs/1411.1308}.

\end{thebibliography}

\appendix

\section{Estimating the variance of unobserved and umodeled parameters} \label{QR}

In this appendix we consider the filtering problem,
\begin{align}\label{a1}
\dot x &= f(x,\theta) \nonumber \\
\dot \theta &= g(\theta,\dot W) \approx \sqrt{Q_{\theta\theta}}\, \dot W  \\
y &= h(x) + \eta \nonumber
\end{align}
where we have access to the noisy observations $y$ at discrete times $t_{i}$ and the noise $\eta$ is normal with variance $R$.  We assume that the model $f$ and the observation function $h$ are known, but that the model $g$ is unknown and we will instead use a white noise model $\tilde g(\theta,\dot W) = \sqrt{Q_{\theta\theta}}\, \dot W$.  We will apply the method of \citep{bs:13} to find the variance $Q_{\theta\theta}$ of the unmodeled parameters $\theta$ in \eqref{a1}.  The method of \citep{bs:13} finds the covariance matrix $Q  = \left( \begin{array}{cc} Q_{xx} & Q_{x\theta} \\ Q_{\theta x} & Q_{\theta\theta} \end{array}\right)$ of the full system $(x,\theta)$.  Thus the technique developed here will allow stochastic forcing with covariance $Q_{xx}$ on the $x$ variable, although none of the examples in this paper have this feature.  

The method of \citep{bs:13} requires us to parameterize the matrix $Q = \sum_{r=1}^R q_rQ_r$ where each $Q_r$ is a fixed matrix of the same size as $Q$ and our goal is to find the parameters $q_r$.  The key to finding the parameter $Q_{\theta\theta}$ for the problem \eqref{a1} is to parameterize the cross covariance $Q_{\theta x}$ between the observed variables $x$ and the unmodeled unobserved variables $\theta$.  Letting $x$ be $n$ dimensional and letting $\theta$ be $m$ dimensional, for each entry $(Q_{\theta x})_{ij}$ we introduce a parameter $q_{in+j}$ which corresponds to a fixed matrix $Q_{i n + j}$ which has a $1$ in the $(i,j)$-th position of $Q_{\theta x}$ and in the $(j,i)$-th position of $Q_{x\theta}$ and is zero otherwise.  Notice, that whenever the matrix $Q= \sum_{r=1}^R q_rQ_r$ is formed, the sub-matrix $Q_{\theta\theta}$ will be identically zero since there are no parameter matrices with nonzero entries in $Q_{\theta\theta}$.  Thus, whenever we form the matrix $Q= \sum_{r=1}^R q_rQ_r$ we must project this matrix onto the nearest symmetric positive definite covariance matrix.  We do this by finding the singular value decomposition $Q = U\Lambda V^\top$ and then replacing $Q$ with $\hat Q = U\Lambda U^\top$.  With this choice of parameterization of $Q$, along with the method of projecting on $\hat Q$, we can apply the method of \citep{bs:13} to find the parameters $q_r$ from an adaptive EnKF.

\end{document}